\documentclass{iopart}

\usepackage{iopams}
\usepackage{graphicx}
\usepackage{subfigure}

\begin{document}

\title[Single photon emission from silicon-vacancy colour centres]{Single photon emission from silicon-vacancy colour centres in CVD-nano-diamonds on iridium}

\author{Elke Neu$^1$, David Steinmetz$^1$, Janine Riedrich-M\"oller$^1$, Stefan Gsell$^2$, Martin Fischer$^2$, Matthias Schreck$^2$ and Christoph Becher$^1$}

\address{$^1$Universit\"at des Saarlandes, Fachrichtung 7.3 (Technische Physik), Campus E2.6, 66123 Saarbr\"ucken, Germany}
\address{$^2$ Universit\"at Augsburg, Lehrstuhl f\"ur Experimentalphysik IV, Universit\"atsstr.1 (Geb\"aude Nord), 86135 Augsburg, Germany  }
\ead{christoph.becher@physik.uni-saarland.de}
\begin{abstract}
We introduce a process for the fabrication of high quality, spatially isolated nano-diamonds on iridium via microwave plasma assisted CVD-growth. We perform spectroscopy of single silicon-vacancy (SiV)-centres produced during the growth of the nano-diamonds. The colour centres exhibit extraordinary narrow zero-phonon-lines down to 0.7 nm at room temperature. Single photon count rates up to 4.8 Mcps at saturation make these SiV-centres the brightest diamond based single photon sources to date. We measure for the first time the fine structure of a single SiV-centre thus confirming the atomic composition of the investigated colour centres.
\end{abstract}

%Uncomment for PACES numbers title message
%\pacs{00.00, 20.00, 42.10}
% Keywords required only for MAST, POB, PM, PM, JOE, JOB?
%\vspace{2pc}
%\noindent{\it Keywords}: Article preparation, IO journals
\submitto{\NJP}
%\tableofcontents
\section{Introduction}
Single photon sources are of great interest for various applications in quantum information, in particular quantum cryptography (e.g.\cite{Gisin2002}). In recent years colour centres in diamond have proven the potential to fulfill the requirements for practical single photon sources such as room temperature operation, photostability and high brightness. Nitrogen-vacancy (NV)-centres \cite{Kurtsiefer2000,Rabeau2007,Beveratos2001}, nickel based NE8 centres \cite{Wu2007,Gaebel2004,Wu2006,Rabeau2005}, nickel-silicon complexes \cite{Steinmetz2010,Aharonovich2009}, chromium related colour centres \cite{Aharonovich2010a,Aharonovich2010b} as well as silicon-vacancy (SiV)-centres \cite{Wang2006} have been employed as single photon sources. The majority of the experiments has been performed on NV-centres as they are easily available in single-crystal diamond \cite{Kurtsiefer2000} and in nano-diamonds \cite{Rabeau2007}, either produced during the growth process or by ion implantation. In addition, due to the extraordinary long spin coherence times, key experiments towards quantum information processing have been realized with NV-centres, e.g. two-qubit quantum-gates \cite{Jelezko2004} or quantum registers based on NV spins \cite{Neumann2010}. However, NV-centres possess the detrimental property of a room temperature emission width of about 100 nm due to strong phonon-coupling.\\
\begin{figure}[h]
\begin{center}
\includegraphics[width=0.6\textwidth]{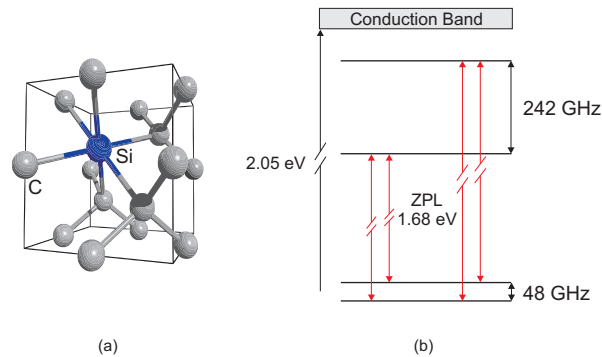}
\caption{(a) Proposed structure of the SiV-centre with a silicon atom in a split-vacancy site (b): Level scheme of the SiV-centre according to \cite{Iakoubovskii2000a,Clark1995}. The optical transitions are not drawn to scale. \label{sivstruct}}
\end{center}
\end{figure}
Silicon-vacancy (SiV-)centres were suggested as an alternative to NV-centres due to their narrow zero-phonon-line (ZPL) width of 5 nm at room temperature, their low phonon-coupling \cite{Wang2006}, as well as their feasible creation by ion implantation or during chemical vapour deposition (CVD)-growth \cite{Zaitsev2001}. A further advantage is their emission in the near infrared at 738 nm in a spectral region where background fluorescence of the surrounding diamond material is weak. First single emitter experiments on SiV-centres created by ion implantation in natural diamonds \cite{Wang2006}, however, revealed unfavourable low single photon emission rates of the order of only 1000 counts/s, despite a short lifetime of 1.2 ns. Radiative quantum yields of only 0.05 were found for SiV-centres in CVD-films \cite{Turukhin1996} and were explained by temperature dependent non-radiative transitions, indicated by a strong increase in the luminescence intensity for temperatures below 100 K \cite{Feng1993}. Furthermore, the luminescence of SiV-centres is quenched by the presence of graphite \cite{Iakoubovskii2000a}, thus defining the need for high purity diamonds.\\
 The structure of the SiV-centre has been intensively investigated: it most probably consists of a silicon atom and a lattice vacancy arranged in a so called 'split-vacancy-configuration', where a substitutional silicon atom relaxes its position towards a neighbouring vacancy \cite{Goss1996} as illustrated in figure \ref{sivstruct}. Photoluminescence excitation measurements suggest a position of the ground state of the centre 2.05 eV below the conduction band edge \cite{Iakoubovskii2000a}. Spectroscopy of high quality single-crystalline diamond at cryogenic temperatures reveals a fine structure of the ensemble ZPL consisting of four lines \cite{Clark1995,Sternschulte1994}. A level scheme including a split excited state with components separated by 242 GHz and a split ground state with components separated by 48 GHz with four allowed optical transitions (see figure \ref{sivstruct}) was proposed to account for the observed fine structure \cite{Clark1995}. This fine structure has never been observed for single SiV-centres so far. \\
In this study we present an advanced material system for single emitter spectroscopy of SiV-centres: We employ a microwave-plasma assisted CVD-growth on nano-diamond seeded iridium substrates to yield high quality nano-diamonds. Single SiV-centres are produced in situ during the CVD-growth due to the presence of silicon impurities. Nano-diamonds exhibit superior properties as hosts for single photon sources as the fluorescence can be efficiently collected due to the lack of total internal reflection at the diamond-air interface \cite{Beveratos2001}. We note that efficient collection of photons can also be achieved using nano-structures in single-crystal diamond, e.g diamond nano-wires \cite{Babinec2010}. A further advantage of nano-diamonds is a low fluorescence background due to the small diamond volume in the excitation laser focus. The iridium substrate provides optimized properties for our application: First of all, the metal substrate exhibits an extremely low fluorescence background. Secondly, spectroscopy on diamond films grown on iridium substrates did not reveal any substrate specific luminescence in our experiments, thus we conclude that growth on iridium does not lead to substrate related colour centres. As a consequence, mostly luminescence-free nano-diamonds are produced. Furthermore, incorporation of the heavy element iridium into the dense crystal lattice of diamond is expected to be disfavoured as compared to many lighter elements (like B, N, Si) due to its high atomic radius, enabling growth of high purity nano-diamonds. Finally, the iridium layer has the potential to alter significantly the radiative properties of the colour centres: in the presence of a metal layer the electromagnetic field of an emitting dipole interacts both with the metal via near field coupling and with the dipole itself via radiation back-action \cite{Buchler2005,Lukosz1977}. The resulting effects include reduction of excited-state lifetimes, enhanced excitation probabilities, as well as enhanced collection efficiency due to radiation pattern modification, and depend critically on the distance of the emitter to the metal layer. Such effects have been investigated in detail for various single emitters e.g. single molecules \cite{Buchler2005} or single CdSe-quantum-dots \cite{Vion2010}.\\
\section{Experimental}
\subsection{Sample Preparation}
\begin{figure}
\begin{center}
\subfigure[\label{seedsizehist}]
{\includegraphics[width=0.4075\textwidth]{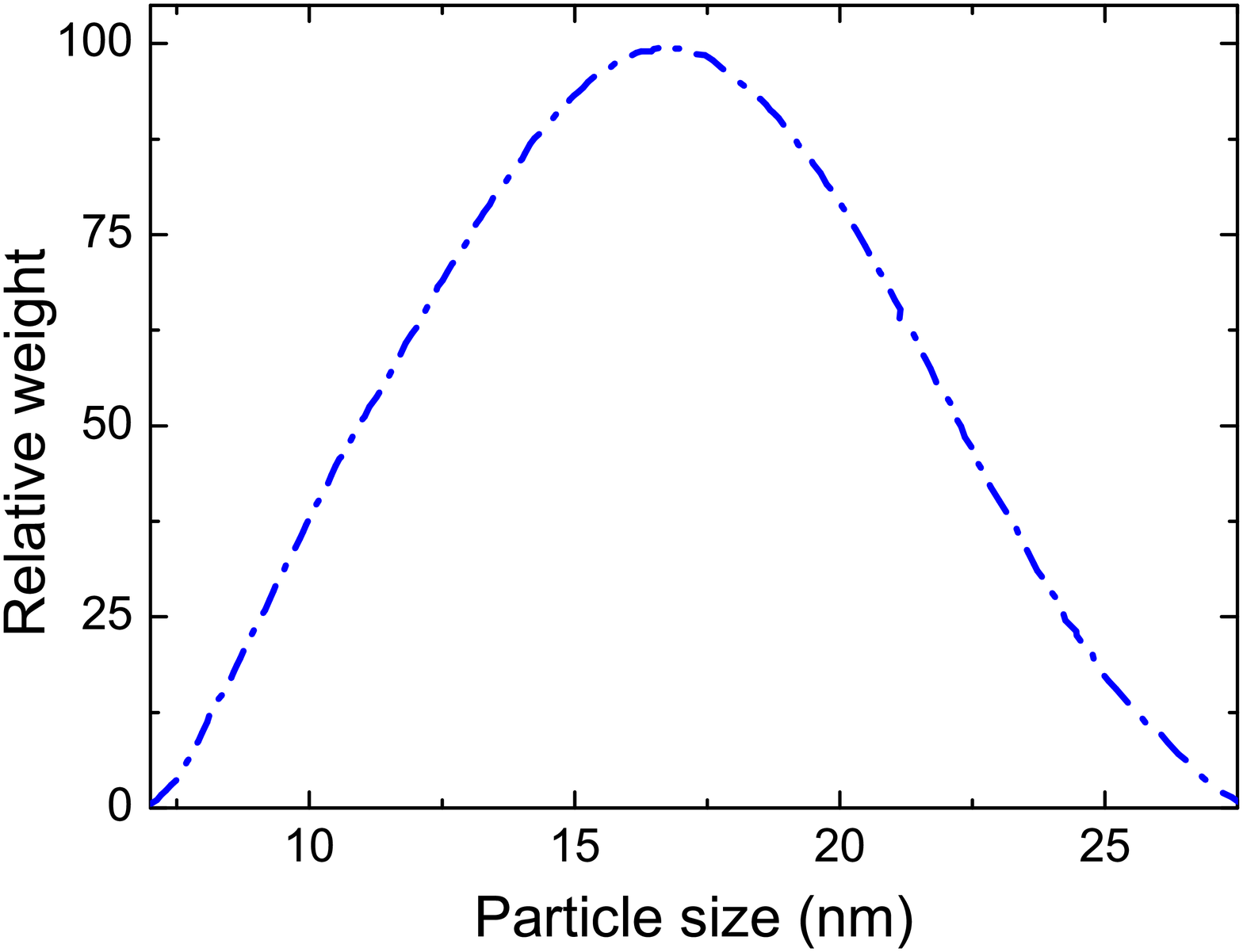}}\hspace{0.01 \textwidth}
\subfigure[\label{NDsizeshist}]
{\includegraphics[width=0.44\textwidth]{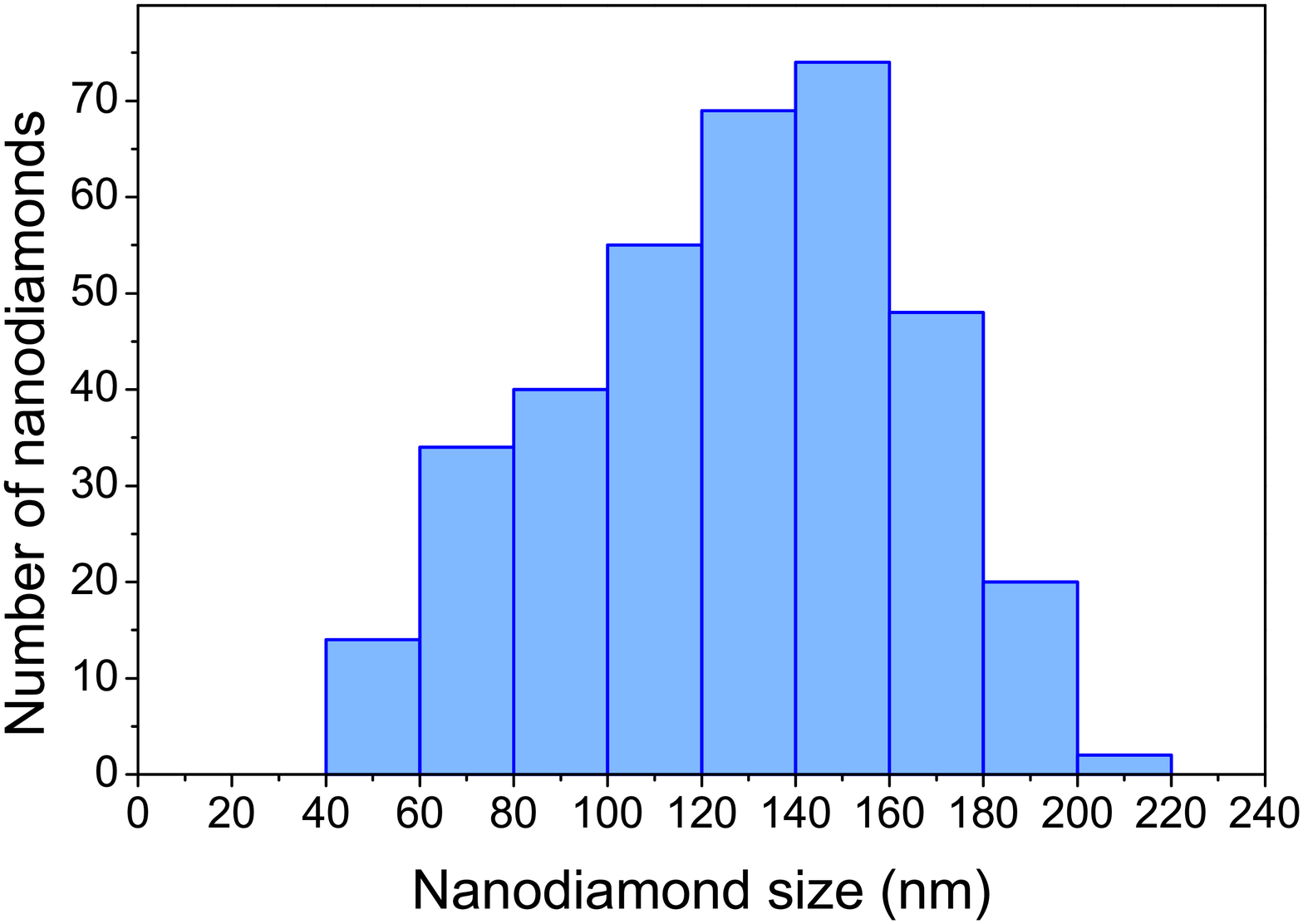}}
\caption{(a) Size distribution of the nano-diamonds used for the seeding process, as measured by the manufacturer. (b) Histogram of nano-diamond sizes after 25 minutes of growth. An overall number of 355 nano-diamonds was evaluated. The overall sample area under investigation was 240 $\mu$m$^2$. The sizes are averaged values, neglecting the different orientations of the diamonds. }
\end{center}
\end{figure}

Nano-diamonds were grown on substrates consisting of a 150 nm thick iridium layer deposited onto a 40 nm thick buffer layer of yttria-stabilized zirconia on silicon \cite{Gsell2004}. For the growth process 10x10 mm$^2$ substrates were used. Prior to the seeding procedure the substrates were cleaned in permonosulfuric acid and rinsed in purified distilled water to remove possible contamination. The seeding procedure employed an aqueous solution of fully de-agglomerated synthetic nano-diamonds (Microdiamant Liquid Diamond MSY) with sizes up to 30 nm which was spin coated onto the substrates. The size distribution of the seed nano-diamonds as measured by the manufacturer is displayed in figure \ref{seedsizehist}: The maximum of the distribution is approximately 17 nm, with a full width at half maximum of about 12 nm. By diluting the solution we are able to conveniently tune the seed density attained by the spin coating. For the investigated samples the solution was diluted to contain approximately 0.6 mg of diamond per litre solution. The seed density estimated from scanning-electron-microscope (SEM) images is roughly 2.5 seeds per $\mu$m$^2$.\\
The seeded substrates were subjected to a microwave plasma assisted chemical vapour deposition process for 25 min. A hydrogen-methane plasma containing 1 \% of methane at a gas pressure of 30 mbar and a microwave power of 2000 W was employed. SiV-centres are created during the CVD-growth due to the presence of silicon impurities originating from the plasma exposure of the silicon substrates at the cut edges of the samples. Regarding the origin of silicon impurities it cannot be excluded that the plasma in contact with the quartz wall of the reactor etches and transfers some additional silicon to the gas phase. However, etching of reactor walls and incorporation of silicon can be controlled to a large extend by the size of the plasma ball or the composition of the gas phase \cite{Ruan1993}. On the other hand, if silicon is used as a substrate we observe the formation of ensembles of SiV-centres in isolated nano-diamonds, similar to the findings of \cite{Stacey2009}, but never single SiV-centres. In this case, the efficient creation of SiV-centres is attributed to the plasma etching of the exposed substrate \cite{Stacey2009,Bergman1993}. Thus, we assume that the silicon substrate is the major source of silicon impurities for the growth on the iridium layer system as well.\\
\begin{figure}[h]
\begin{center}
\includegraphics[width=0.6\textwidth]{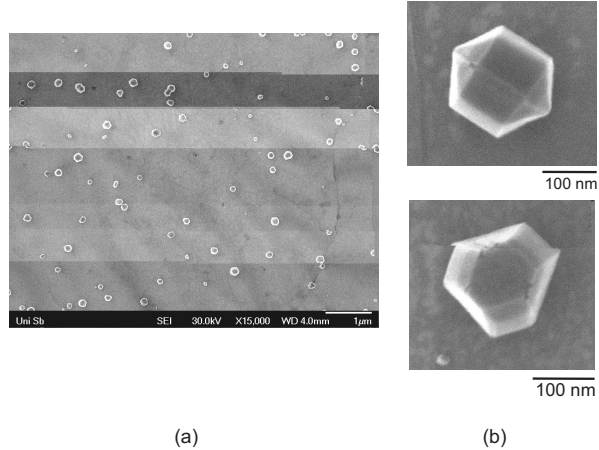}
\caption{Scanning-electron-microscope images of CVD-nano-diamonds. (a) Overview image area 8x6 $\mu$m$^2$, (b) detailed images of typical well faceted nano-diamonds \label{NDSEMpics}}
\end{center}
\end{figure}
Morphological characterization of the nano-diamonds was performed using a high resolution SEM (JEOL JSM-7000 F). Figure \ref{NDSEMpics}(a) displays an overview of an 8x6 $\mu$m$^2$ area of the sample, confirming the growth of isolated nano-diamonds with a density of 1.7-2.2 CVD-nano-crystals per $\mu$m$^2$, slightly lower than the estimated seed density. Detailed images (figure \ref{NDSEMpics} (b)) reveal well faceted nano-diamonds. The nano-diamonds show various crystal orientations, reproducing the different orientations of the seed crystals \cite{Stacey2009}. In principle however, the iridium layers provide an additional option due to their unique properties as heteroepitaxy substrate for diamond(001) \cite{Gsell2004} and diamond(111) films \cite{Fischer2008}: Synthesis of epitaxial diamond nano-crystals on iridium would yield SiV-centres residing in a well-aligned diamond matrix, giving rise to a reproducible and well-defined orientation of the colour centre dipole.\\
By analyzing the SEM images we obtain the size distribution of the nano-diamonds displayed in figure \ref{NDsizeshist}. We identify a mean size of the crystals of 130 nm with a standard deviation of 40 nm. The distribution is slightly asymmetric, decaying more slowly towards small crystal sizes, thus qualitatively resembling the size distribution of the seed diamonds.\\
The measurements described in the following were performed without any post-growth treatment of the nano-diamonds. With the given density of nano-crystals roughly one bright SiV-centre is observed in 50x50 $\mu$m$^2$ thus enabling feasible single emitter characterization.
\subsection{Experimental Setup}
To analyze the nano-diamond fluorescence we use a confocal microscope setup. As excitation source a 671 nm frequency doubled DPSS laser focused through a microscope objective with NA=0.8 is employed. By using a dichroic beamsplitter and dielectric filters the fluorescence in the spectral range of 730-750 nm is separated from reflected laser light and coupled into a multimode fibre simultaneously serving as pinhole for the confocal setup. The fluorescence is directed to a Hanbury-Brown-Twiss setup (HBT) employing two avalanche photodiodes (APDs) (Perkin Elmer SPCM-AQRH-14) to measure the intensity autocorrelation g$^{2}$ of the emitted photons.\\
For spectral characterization the fluorescence is analyzed by a grating spectrometer (Horiba Jobin Yvon, iHR 550). For room temperature (cryogenic temperature) spectroscopy we employ a grating with 600 (1800) grooves/mm, yielding a resolution of approx. 0.18 (0.06) nm, respectively. Taking into account the quantum efficiency of the APDs (typically $70\%$), the solid angle fraction corresponding to the NA of the objective and the transmission of the optical components we estimate an overall detection efficiency of $\eta=5.5\%$.
To facilitate single emitter characterization at cryogenic temperatures a flow cryostat (Janis Research, ST-500LN) operated with liquid helium is used.
\section{Results}
\subsection{Single emitter spectroscopy at room temperature}
\subsubsection{Emission spectra}
\begin{figure}[h]
\begin{center}
\includegraphics[width=0.4\textwidth]{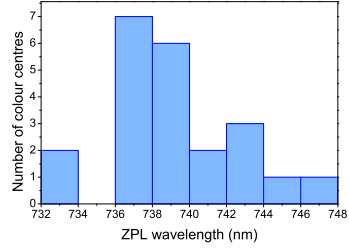}
\caption{Histogram of the observed ZPL wavelengths of 22 colour centres. For the histogram also nano-diamonds with multiple colour centres were taken into account.   \label{histemitters}}
\end{center}
\end{figure}

Confocal scans of the samples reveal nano-diamonds with bright luminescence in the spectral range 730-750 nm. A histogram of the observed lines is shown in figure \ref{histemitters}. For the histogram crystals containing single and multiple colour centres were taken into account. The measured line positions spread from 732 nm to 748 nm, although the majority of the observed colour centres shows a ZPL between 736 and 740 nm. From investigations of poly-crystalline diamond films, ensembles of SiV-centres are known to exhibit a room temperature line-width of 6 nm  with a temperature independent inhomogeneous line-width of up to 4.4 nm \cite{Gorokhovsky1995}, depending on film quality. Thus the inhomogeneous broadening of the SiV-centre emission explains most of the observed spread of the line positions.\\
\begin{figure}
\begin{center}
\subfigure[\label{RTZPL}]
{\includegraphics[width=0.49\textwidth]{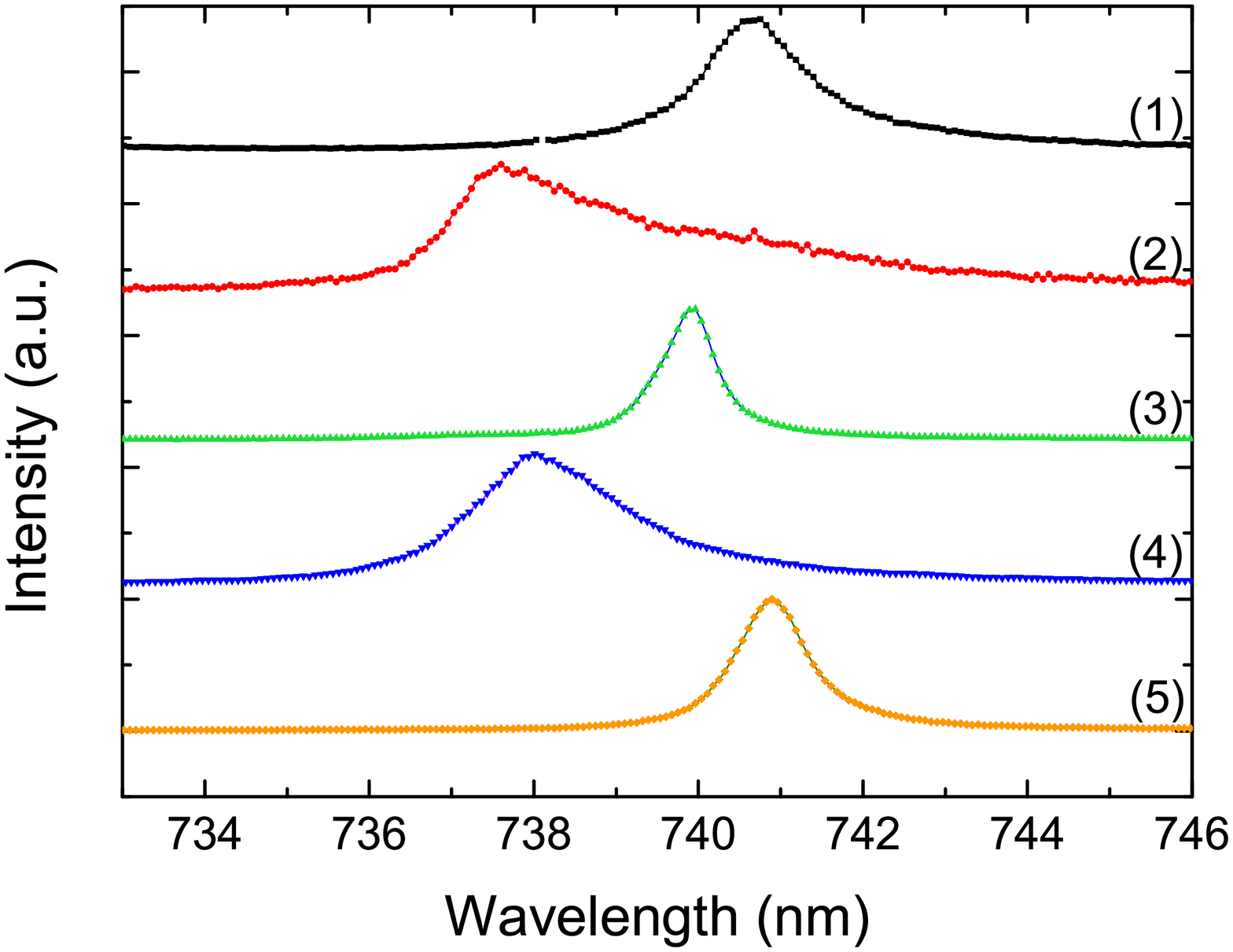}}
\subfigure[\label{RTsidebands}]{
\includegraphics[width=0.49\textwidth]{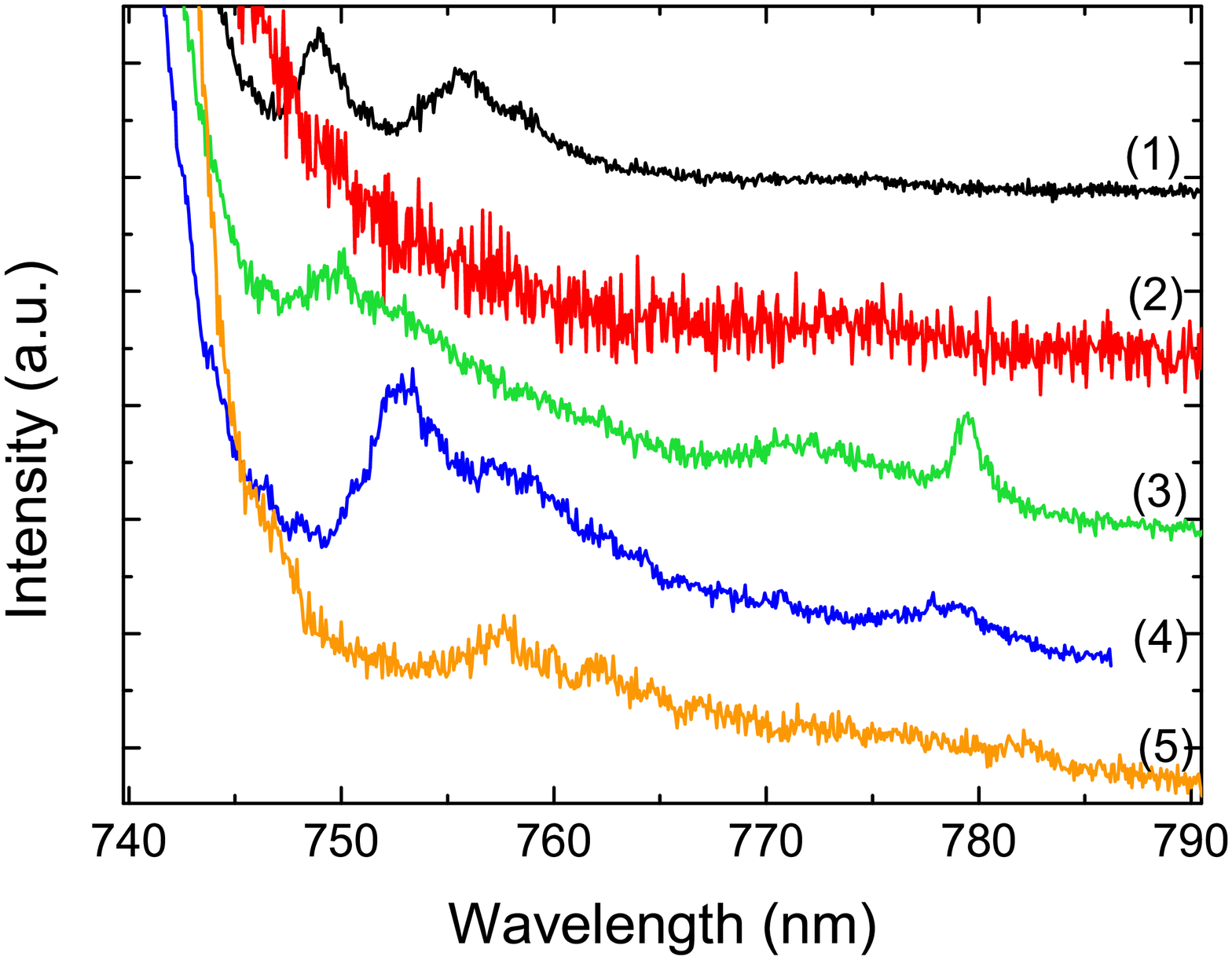}}
\caption{Room temperature spectra of the five colour centres under investigation. (a) ZPLs of the colour centres. The spectra have been normalized and displaced for clarity. (b) Sideband structure of the emitters. Spectra have been scaled and displaced to give a clear impression of sideband structure. }
\end{center}
\end{figure}
In the following we present a detailed characterization of five different single colour centres in individual nano-diamonds. Figure \ref{RTZPL} displays spectra of the ZPLs of these centres. The peak positions are ranging from 737.6 nm to 740.9 nm. Emitter (2) exhibits an asymmetric line shape, which is fitted best by introducing three lines at the positions listed in table \ref{overviewempar}. Remarkably, all colour centres display a room temperature ZPL line-width of less than 2.2 nm. Emitter (3) shows a line-width of only 0.7 nm, which is to our knowledge the narrowest room temperature line-width measured for a diamond colour centre up to now. Thus the line-widths obtained are comparable to NE8 centres (1.2 nm \cite{Gaebel2004}, 1.5nm \cite{Rabeau2005}) delivering the narrowest line-widths so far and superior to the line-widths observed for single SiV-centres in former studies (5-7 nm \cite{Wang2006,Wang2007b}). The observed narrowing of the ZPLs compared to former studies might be related to the broadening mechanisms involved: as described e.g. in \cite{Davies1981} the width of the ZPL at room temperature is determined by homogeneous broadening due to electron-phonon scattering induced by a quadratic electron-phonon interaction. This broadening relates to the phonon density of states which critically depends on local lattice distortions. Such local variations are visible e.g. in varying phonon-sideband spectra as discussed below. Thus we assume that the high crystal quality of our nano-diamonds and a low impurity content might be responsible for the narrow ZPLs observed.\\
\begin{table}
\caption{\label{overviewempar}Overview of the luminescence parameters for the five different colour centres presented in this publication labelled emitter (1) to (5). $\lambda_{Peak}$ is the peak wavelength of the ZPL and $\Delta \lambda$ its width. P$_{Sat}$ gives the saturation power from fitting equation (\ref{Satfunc}), I$_{sat}$ the respective saturation intensity. I$_{\infty}$ denotes the single photon count rate at saturation. Finally the Debye-Waller-Factors (DW) and the Huang-Rhys-Factors (S) are listed. }
%\begin{indented}
\lineup
%\item[]
\begin{tabular}{@{}llllllll}
\br
Emitter&$\lambda_{Peak}$ (nm)&$\Delta \lambda$ (nm)&P$_{Sat}$ ($\mu$W) &I$_{Sat}(\frac{kW}{cm^2})$&I$_{\infty}$(kcps)& DW& HR\cr
\mr
(1)&740.7& 1.6  &\014.3&\01.2&\0263&0.76&0.28\cr
(2)&737.6&1.2&\043.0&\03.7&\0512&0.84&0.18\cr
{}&738.7&2.0&&&&\cr
{}&740.8&4.9&&&&\cr
(3)&739.9 &0.7&\040.9&\03.5&\0395&0.79&0.24\cr
(4)&738.2&2.2 &306.7&26.1&4828&0.75&0.29\cr
(5)&740.9 &1.0 &\073.4&\06.2&2395&0.88&0.13\cr
\br
\end{tabular}
%\end{indented}
\end{table}
Another important issue for the use of colour centres as single photon sources is that the strength of the electron-phonon-coupling determines both the fraction of the emission into the ZPL and the overall emission bandwidth including the phonon-sidebands. The phonon-coupling is measured either by the Debye-Waller-Factor (DW) or the Huang-Rhys-Factor (S). The Debye-Waller-Factor is defined as the integrated luminescence intensity of the ZPL I$_{zpl}$ divided by the integrated luminescence intensity of the colour centre I$_{tot}$ \cite{Gaebel2004}. The Huang-Rhys-Factor S is defined by $\frac{I_{zpl}}{I_{tot}}=\exp(-S)$ \cite{Walker1979}. As listed in table \ref{overviewempar} emitters (1), (3) and (4) show Debye-Waller-Factors of 0.75 to 0.79 corresponding to Huang-Rhys-Factors of 0.29 to 0.24, thus corresponding very well to $S=0.24$ determined for ensembles of SiV-centres \cite{Zaitsev2001}. Emitters (2) and (5) exhibit further reduced phonon-coupling resulting in Debye-Waller-Factors of 0.84 and 0.88 respectively. The structure of the vibronic sidebands of the emitters is depicted in figure \ref{RTsidebands}. Table \ref{sidebandsempar} summarizes the energies of the observed features with respect to the ZPLs. The sidebands measured here resemble sidebands previously observed for SiV-ensembles, especially around 38-42 meV, 65 meV and 83-85 meV, as indicated in table \ref{sidebandsempar}.
\begin{table}[b]
\caption{\label{sidebandsempar}Overview of sideband features observed for the different emitters. For comparison sidebands observed in former work are listed. All energies given in meV.  }
%\begin{indented}
\lineup
%\item[]
\begin{tabular}{@{}lllllllllll}
\br
Emitter & S$_{obs}$ & S$_{Lit}$ & S$_{obs}$ & S$_{Lit}$ & S$_{obs}$ & S$_{Lit}$ & S$_{obs}$ & S$_{Lit}$ & S$_{obs}$ & S$_{Lit}$ \cr
\mr
(1)& 18.3 &-&33.7&38.0 \cite{Sittas1996}& - &-& - &-&-&-\cr
(2)& - &-&-&-&- &-&- &-&79.2&83.0 \cite{Sittas1996} \cr
(3)& 22.6 &-&- &-&- &-&69.6&65.8 \cite{Gorokhovsky1995}&84.1& 85.0 \cite{Sternschulte1994}\cr
(4)& - &-&32.8 &38.0 \cite{Sittas1996}&43.9&42.3 \cite{Gorokhovsky1995}&- &-&88.0&85.0 \cite{Sternschulte1994}\cr
(5)& - &-&36.9 &38.0 \cite{Sittas1996}&46.8&42.3 \cite{Gorokhovsky1995}&- &-&88.4&85.0 \cite{Sternschulte1994}\cr
\br
\end{tabular}
%\end{indented}
\end{table}Despite these similarities the vibronic sideband spectra of the individual crystals differ significantly. A related effect has been reported by Gorokhovsky et al. for SiV ensembles in poly-crystalline CVD-films \cite{Gorokhovsky1995}: only by resonant excitation selecting a narrow sub-ensemble a clear sideband structure was observed, while it was washed out for off-resonant excitation due to inhomogeneous effects in the different crystals. Similar considerations are relevant for isolated nano-diamonds taking into account size effects and stress in the individual nano-diamonds.\\
\subsubsection{Single photon count rates \label{photoncount}}
\begin{figure}[h]
\begin{center}
\subfigure[\label{satem13}]
{\includegraphics[width=0.49\textwidth]{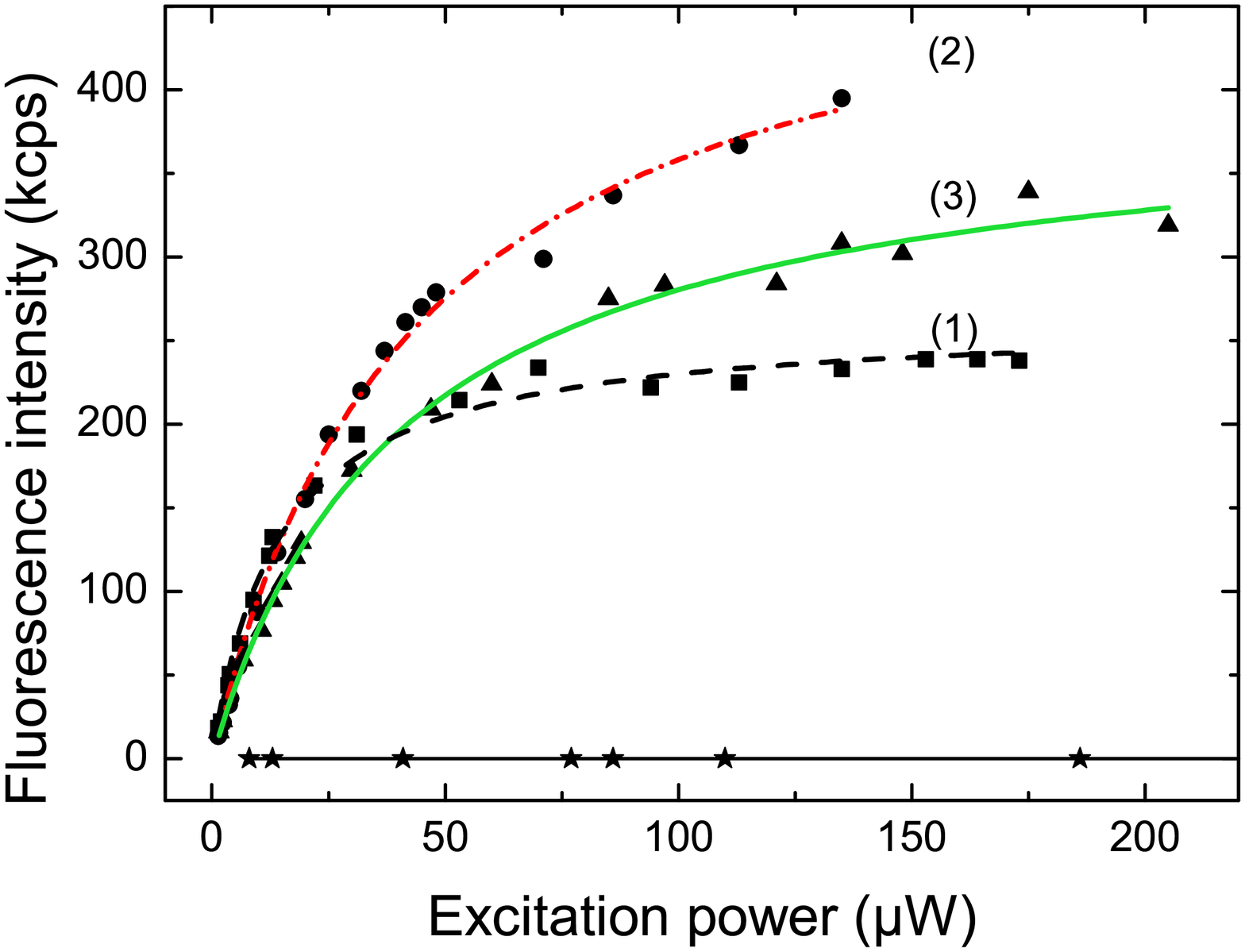}}
\subfigure[\label{satem45}]{
\includegraphics[width=0.49\textwidth]{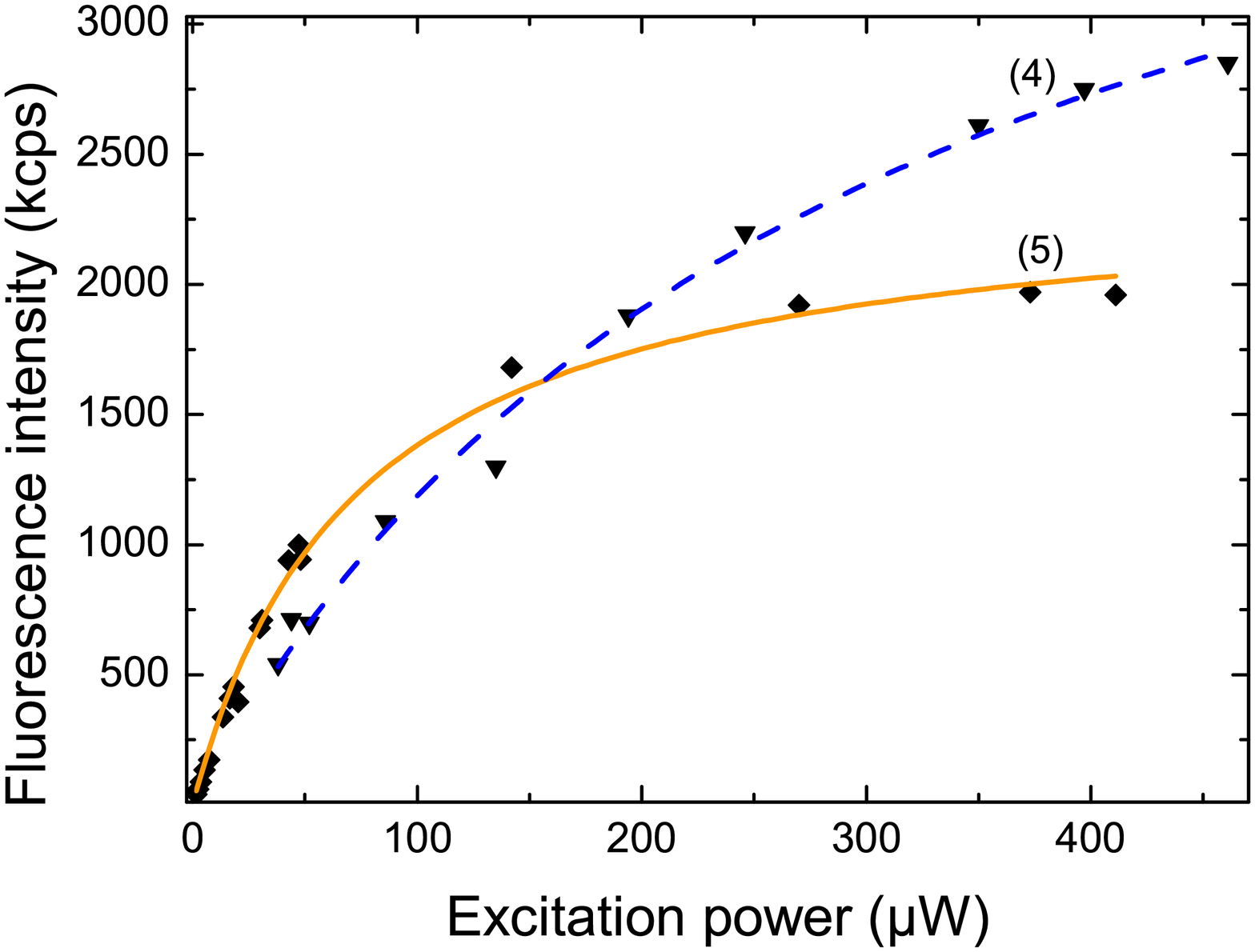}}
\caption{Single photon count rates of the five colour centres under investigation, recorded as a function of excitation power. (a) Count rates of emitters (1)-(3), (b) count rate of emitters (4) and (5). The lines display the fitted functions according to equation (\ref{Satfunc}). The parameters obtained from the fitted functions are summarized in table \ref{overviewempar}. In figure \ref{satem13} the typical background level recorded on the substrate is displayed (filled stars).  \label{sat}}
\end{center}
\end{figure}

As an important figure of merit for a single photon source we investigate the photon count rate obtained at saturation. Figure \ref{sat} displays the saturation curves for the five emitters under investigation. The count rate I in dependence of the excitation power P has been fitted with the function
\begin{equation}
      \textrm{I}=\textrm{I}_\infty\frac{\textrm{P}}{\textrm{P}+\textrm{P}_{sat}}.
      \label{Satfunc}
\end{equation}
Fitting parameters are the saturation power $\textrm{P}_{sat}$ and the maximum obtainable count rate $\textrm{I}_{\infty}$. The values determined from the fitted curves are summarized in table \ref{overviewempar}. By taking into account the transmission of the laser light through the microscope objective and the spot size of the focus ($\frac{1}{e^2}$ radius) of approx. 0.51 $\mu$m, we estimate the intensity impinging on the emitters. The results are also given in table \ref{overviewempar}.\\
Emitter (2) exhibits a remarkably low saturation power of only 14.3 $\mu$W, indicating a very efficient excitation path. In former studies \cite{Wang2006} a saturation power of 6.9 mW for an excitation wavelength of 685 nm was determined. The very efficient excitation observed here with saturation powers below 310 $\mu$W for all emitters allows for a significant reduction of the background fluorescence. In figure \ref{satem13} the typical background level on the substrate 2 $\mu$m away from a luminescent nano-diamond is displayed. Even for the highest excitation powers a background of only 1.5 kcps is observed. Thus, the background fluorescence from the substrate is at least 175 times lower than typical fluorescence rates observed, demonstrating a major benefit of the iridium substrates. Note that this does not give reliable information on the signal to background ratio obtained for the emitters: the background due to fluorescence of the nano-crystal cannot be directly measured as it cannot be separated from the colour centre luminescence. Nevertheless the properties of the measured g$^2$ functions confirm very high signal to background ratios as discussed in section \ref{g2modelling}.\\  The observed photon count rates show a large variation: Emitters (1), (2) and (3) display maximum count rates between 263 and 512 kcps, whereas emitters (5) and (4) feature maximum count rates of 2.4 and 4.8 Mcps. The latter rate, to our knowledge, is the highest photon count rate observed for a single colour centre in diamond up to now, surpassing recently investigated single photon sources based on chromium colour centres delivering a maximum of 3.2 Mcps \cite{Aharonovich2010a}. Furthermore, the count rate observed here is more than three orders of magnitude higher than the count rates obtained in former studies on SiV-centres in single-crystal diamond (approx. 1000 cps, \cite{Wang2006}). The enhancement compared to the count rates for SiV-centres in bulk diamond is not exclusively explained by the enhanced fluorescence extraction efficiency from the nano-crystals compared to bulk diamond which has been measured to enhance the fluorescence rate by a factor of 11 \cite{Wang2007b}. Moreover, the good crystalline quality of the nano-diamonds as observed in SEM-investigations might lead to enhanced fluorescence properties if non-radiative processes are suppressed. Additionally, the presence of the iridium layer influences the fluorescence properties by narrowing the emission cone of the colour centres thus enhancing the collection efficiency \cite{Buchler2005,Vion2010}. The iridium layer might also lead to a weak enhancement of the radiative decay rates of the colour centres as well as a significant enhancement of the non-radiative decay rates due to near field coupling and absorption (see section \ref{g2modelling}). As these effects depend critically on the distance of the emitting dipole from the metal layer they vary with the unknown position of different emitting colour centres inside our nano-diamonds of around 130 nm size, thus contributing to the spread in count rates. In addition, the undefined dipole orientation of the colour centres gives rise to changes in the collection efficiency: for the microscope objective employed (NA=0.8) the collection efficiency varies by a factor of 2.4 for dipoles oriented perpendicular and parallel to the optical axis of the microscope objective \cite{Plakhotnik1995}.  Finally, as discussed in section \ref{g2modelling} the observed colour centres also exhibit different population dynamics including the participation of shelving states leading to different brightness.\\ The investigated emitters did not show blinking, nevertheless emitter (4) was permanently photo-bleached after approximately one and a half hour of observation, when applying high excitation power to perform the saturation measurement. If a change in the charge state was responsible for the bleaching, enhanced stability might be gained by controlling the surface termination of the nano-diamonds as shown for NV-centres \cite{Fu2010}. Another route might be enlarging the nano-diamond size to reduce surface effects. These approaches are currently under investigation.\\
\begin{figure}[h]
\begin{center}
\subfigure[\label{excpol}]
{\includegraphics[width=0.49\textwidth]{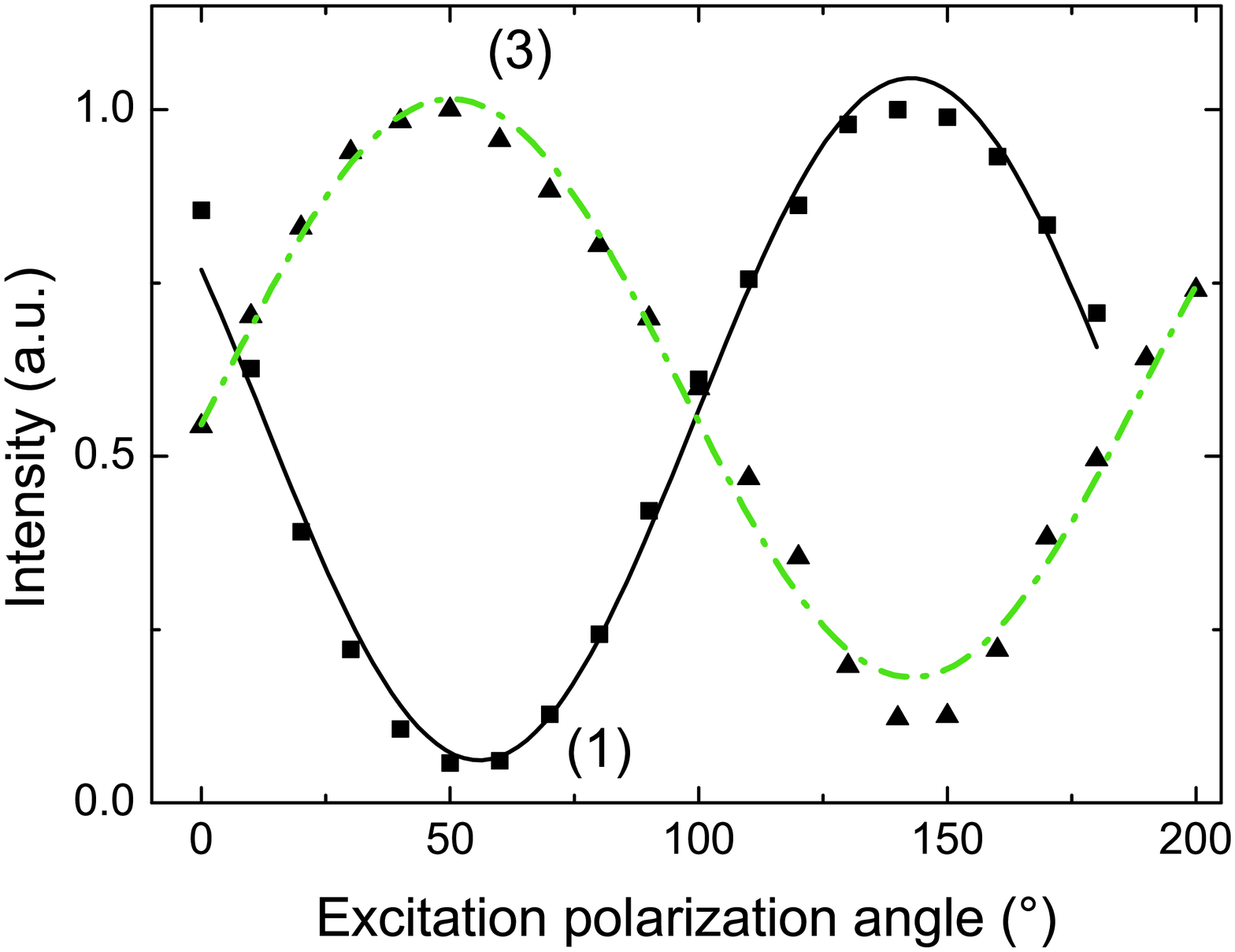}}
\subfigure[\label{plpol}]{
\includegraphics[width=0.49\textwidth]{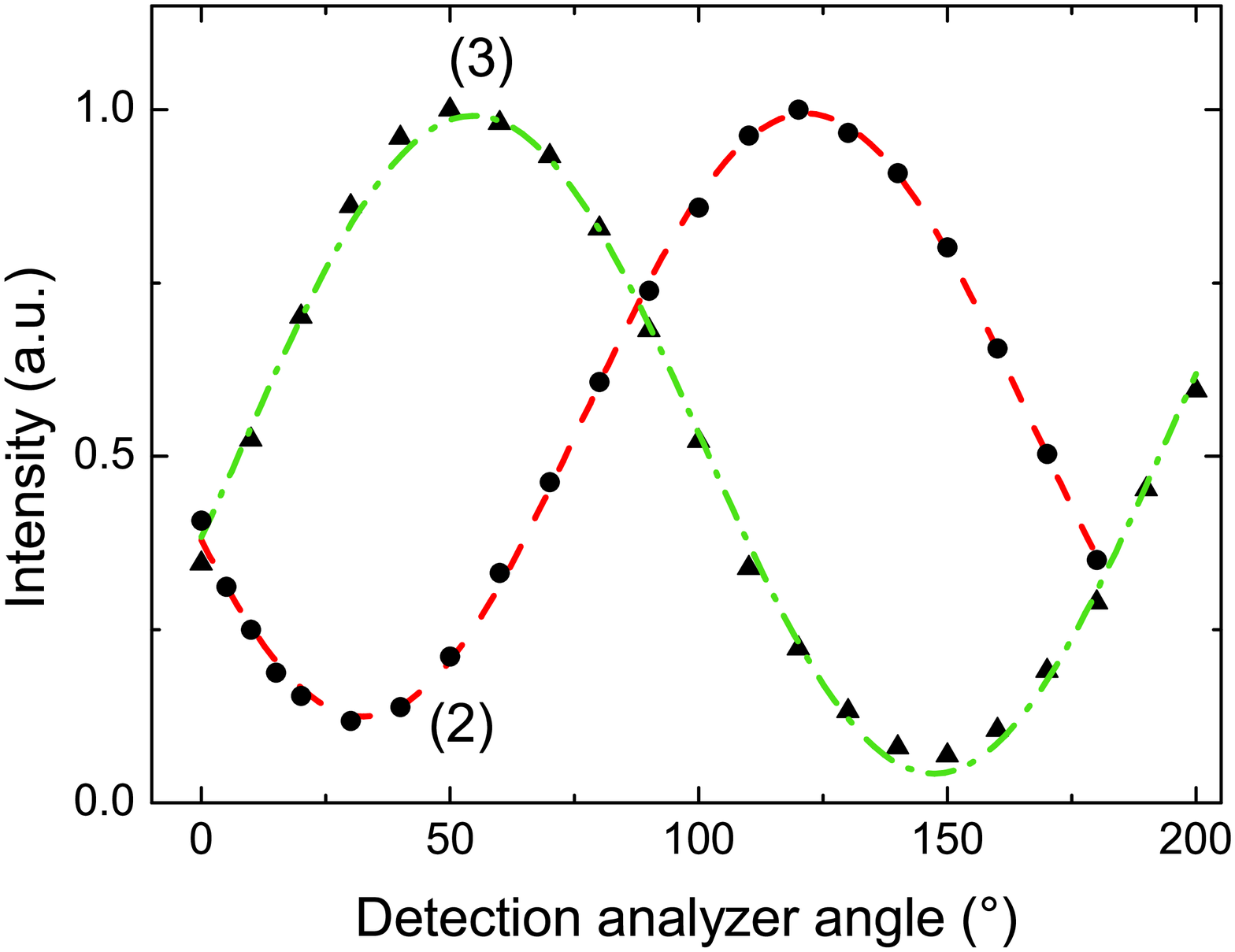}}
\caption{Polarization dependent count rates of selected emitters (a) as a function of the excitation polarization (b) for a fixed excitation polarization in dependence of the analyzer angle. Note that the position of the minima and maxima shifts for different emitters due to the varying orientation of the colour centres in the nano-crystals.\label{Polem} }
\end{center}
\end{figure}
\subsubsection{Polarization measurements}
The polarization properties of the colour centres play a crucial role in their use as single photon sources: if the colour centre preferentially absorbs linearly polarized light one can optimize the excitation by choosing the fitting linear polarization, thus gaining signal to background ratio. Linearly polarized emission is advantageous for applications of single photon sources, e.g. in quantum cryptography (BB84-protocol), which require photons of a defined polarization. To measure the polarization properties of the colour centres in absorption a half wave plate is used to rotate the polarization angle of the excitation laser light. To determine the polarization properties of the emitted photons we use a linear polarization analyzer in the detection path. Figure \ref{Polem} displays the curves measured for selected emitters. The intensity $I$ in dependence of the polarizer angle is fitted using a sine-square function. To describe the contrast obtained we calculate the visibility $V$
\begin{equation}
      V=\frac{I_{max}-I_{min}}{I_{max}+I_{min}}\,.
\end{equation}
\begin{table}[h]
\caption{\label{visibil}Visibility of the polarization dependent intensity. V$_{exc}$ denotes the visibility obtained from rotating the excitation polarization. V$_{pl}$ denoted the visibility obtained by rotating a linear polarization analyzer in the detection path, while fixing the excitation polarization. For emitter (4) the polarization of the luminescence has not been measured.    }
\begin{indented}
\lineup
\item[]\begin{tabular}{lll}
\br
Emitter&V$_{exc} (\%)$ & V$_{pl} (\%)$\cr
\mr
1&88.9 & 87.6\cr
2&83.0 & 77.7\cr
3&78.0 & 91.0\cr
4&84.9 & --\cr
5&70.0 & 84.3\cr
\br
\end{tabular}
\end{indented}
\end{table}
 Table \ref{visibil} summarizes the visibilities obtained from the measurements. If the colour centre can be described as a single dipole in absorption we expect a visibility of close to $100\%$ \cite{Ha1999} with a change from maximum to minimum intensity corresponding to a change of polarization angle of 90$^{\circ}$. This behaviour is basically observed for emitters (1), (2) and (4), although the contrast does not reach $100\%$. The deviation is mainly due to polarization changes induced by the dichroic beam splitter: for s and p-polarization the dichroic mirror does not change the polarization of the laser light, whereas for light polarized at 45$^{\circ}$ we observe a loss of linear polarization degree of $10\%$. Thus the depolarization due to the beam splitter might give an explanation for the reduced visibility for emitters (1), (2) and (4), yielding evidence for single dipole behaviour in excitation. However, as the visibility amounts only to $70\%$ and $78\%$ for emitters (3) and (5) the contribution of a second dipole to the excitation as e.g observed for the NV-centre \cite{Alegre2007} cannot be excluded.\\  For the emitted fluorescence similar experimental considerations hold: when passing the dichroic mirror
 polarization angle dependent rotation and loss of linear polarization occur. These effects have been qualitatively observed for reflected laser light. Further investigations have to reveal whether the loss in linear polarization degree is due to an elliptical polarization as stated in \cite{Wu2006}. Thus, due to depolarizing effects, we do not expect to observe $100\%$ visibility even for fully linear polarized emission.
 Furthermore, as a consequence of the angle dependent polarization rotation effects, the polarization angle measured behind the dichroic mirror differs from the polarization angle of the emitted light.
 In addition to the depolarizing effects in the detection path, we have to take into account a loss of polarization contrast due to imaging with a high NA objective \cite{Fourkas2001}: the polarization visibility of a single dipole depends critically on its orientation: for a dipole with its axis perpendicular to the optical axis of the microscope  $100\%$ visibility is theoretically obtained. On the other hand, for a dipole with its axis parallel to the optical axis of the microscope objective no net contrast is obtained as the polarization is not constant over the collimated fluorescence beam ('polarization anisotropy'). As the orientation of the nano-diamonds on the sample is random, the orientation of individual SiV-centre dipoles is random as well. Thus, we are in principle not able to determine whether a polarization contrast deviating from $100\%$ is due to the orientation dependent loss of contrast or due to a possible contribution of a second dipole moment.\\
 In our experiments on polarization contrast of SiV-fluorescence  three emitters (1), (3) and (5) show high visibilities of $87.6\%$,  $91\%$ and  $84.3\%$, respectively. On basis of the experimental considerations described above we can assume that these emitters show close to $100\%$ visibility of the emitted light, thus evidencing single dipoles oriented approximately perpendicular to the optical axis of the setup. The measured visibility of emitter (2) amounts only to $77.7\%$. Here we cannot distinguish whether the dipole is oriented at an angle between 0 and 90$^\circ$ to the optical axis or a second dipole contributes to the transition. To our knowledge the nature of the electronic states of the SiV-centre is still subject to debate \cite{Clark1995,Goss1996} and no investigations on dipoles contributing to transitions of the SiV-centre are reported in literature. Thus this topic needs further investigation. \\
\subsubsection{Photon correlation measurements and theoretical modelling \label{g2modelling}}
\begin{figure}
\begin{center}
\subfigure[\label{g2em1}]
{\includegraphics[width=0.45\textwidth]{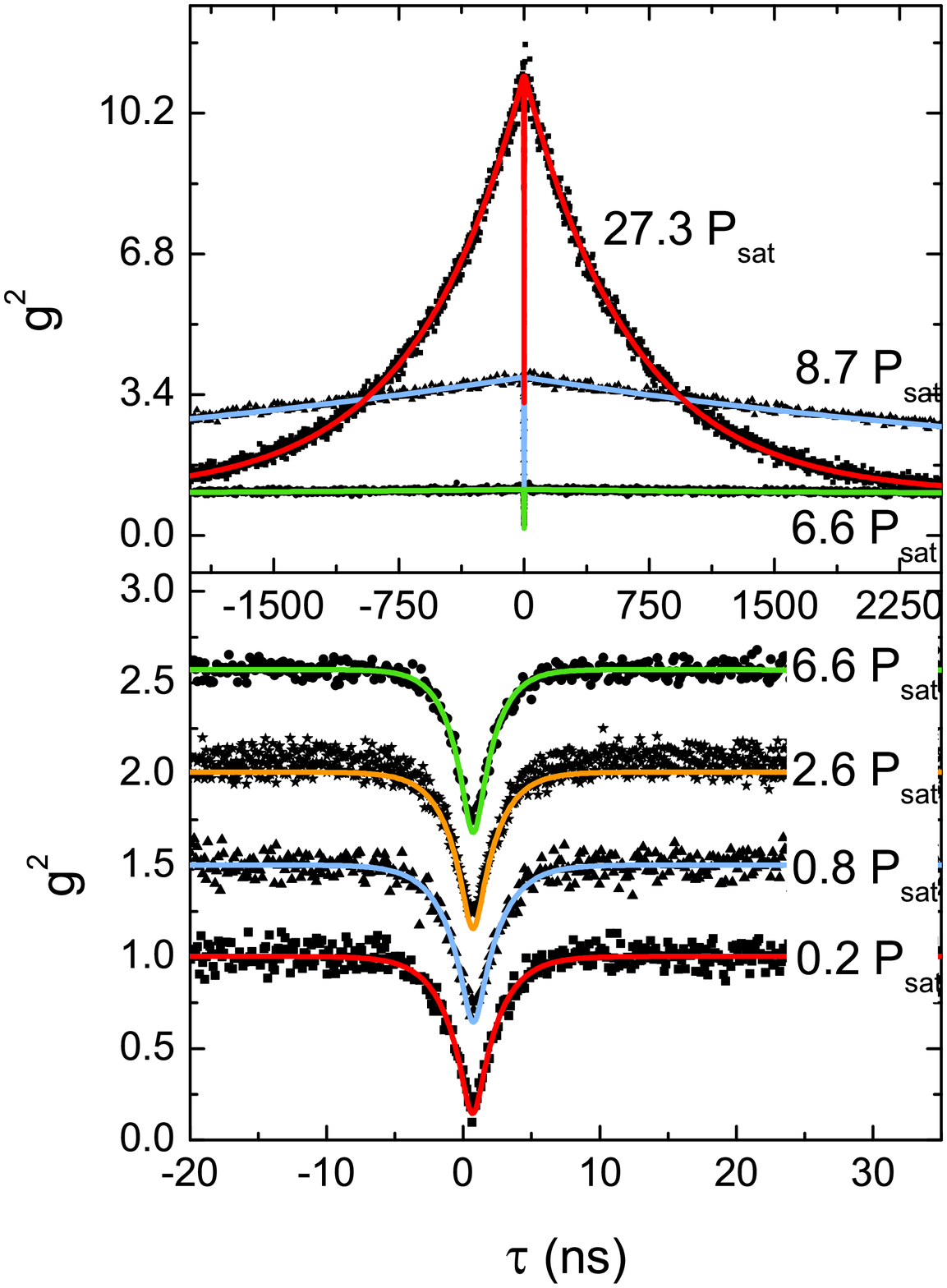}}
\subfigure[\label{konstem1}]{
\includegraphics[width=0.45\textwidth]{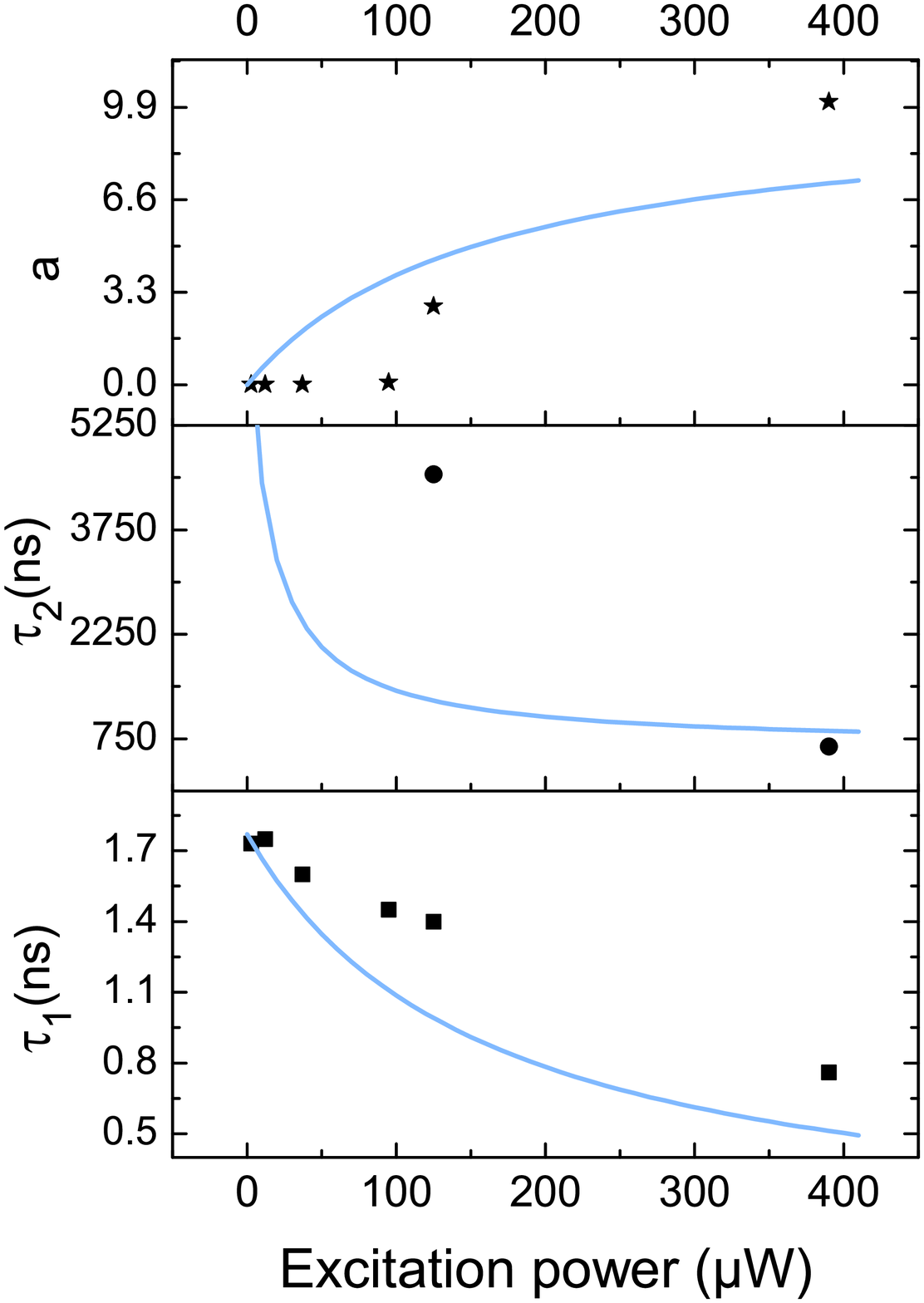}}
\caption{Population dynamics of emitter (1). (a) g$^2$-functions for different excitation powers, given in units of the saturation power P$_{Sat}=14.3$ $\mu$W. For lower excitation power the region of short delays is plotted (lower graph), while for higher powers a larger region of delays is plotted, to permit observation of the bunching behaviour. Note that for low power consecutive g$^2$ functions have been displaced by 0.5 for clarity. Fits confirm that the residual g$^2(0)$ values are due to the timing jitter only, unambiguously proving single emitter behaviour. (b) constants $\tau_1$,$\tau_2$ and $a$ obtained from the g$^2$ functions. $\tau_2$ constants for $\textrm{P}<125\mu\textrm{W}$ are not given as they could not be reliably determined due to the very low value of $a$.    \label{em1dynamics}}
\end{center}
\end{figure}

\begin{figure}
\begin{center}
\subfigure[\label{g2em3}]
{\includegraphics[width=0.45\textwidth]{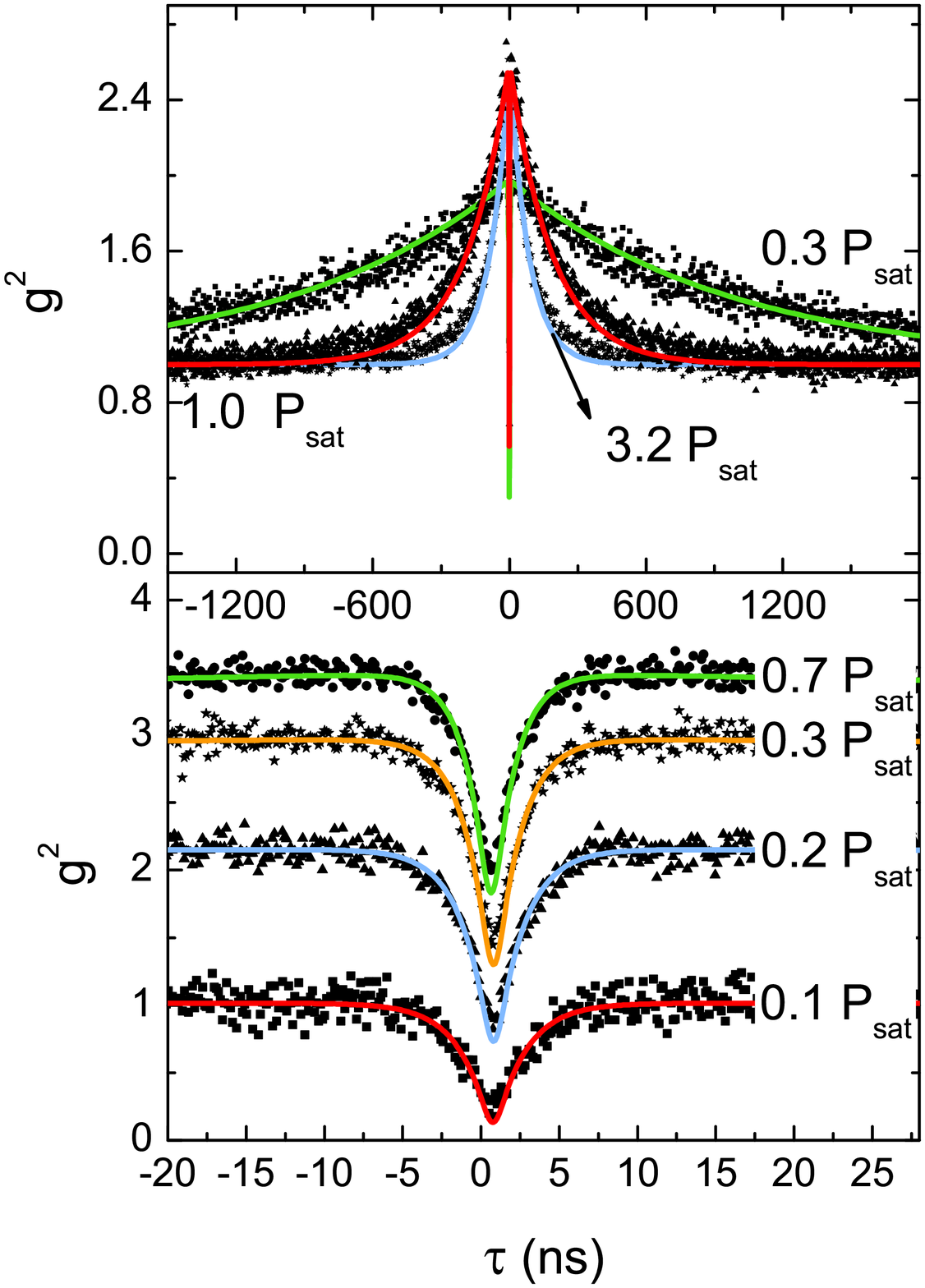}}
\subfigure[\label{konstem3}]{
\includegraphics[width=0.45\textwidth]{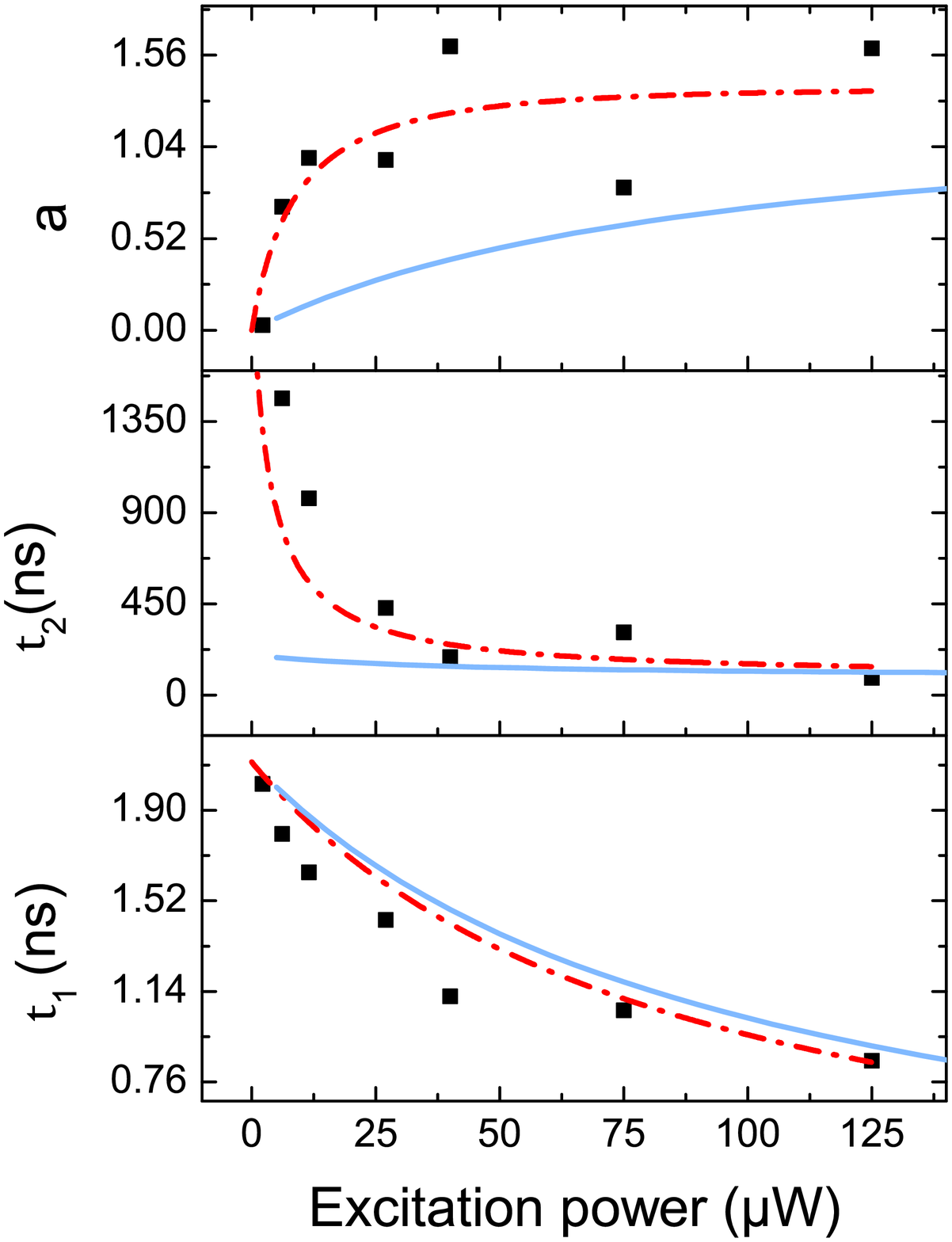}}
\caption{Population dynamics of emitter (3). (a) g$^2$-functions for different excitation powers, given in units of the saturation power P$_{Sat}=40.9$ $\mu$W. For lower excitation power the region of short delays is plotted (lower graph), while for higher powers a larger region of delays is plotted, to permit observation of the bunching behaviour. Note that for low power consecutive g$^2$ functions have been displaced by 0.5 for clarity. Fits confirm that the residual g$^2(0)$ values are due to the timing jitter only, unambiguously proving single emitter behaviour. (b) constants $\tau_1$,$\tau_2$ and $a$ obtained from the g$^2$ functions. solid blue lines: intensity dependence obtained from the simple three level model, dashed red lines: intensity dependence obtained from the extended model including an intensity dependent de-shelving  \label{em3dynamics}}
\end{center}
\end{figure}

\begin{figure}
\begin{center}
\subfigure[\label{g2em2}]
{\includegraphics[width=0.45\textwidth]{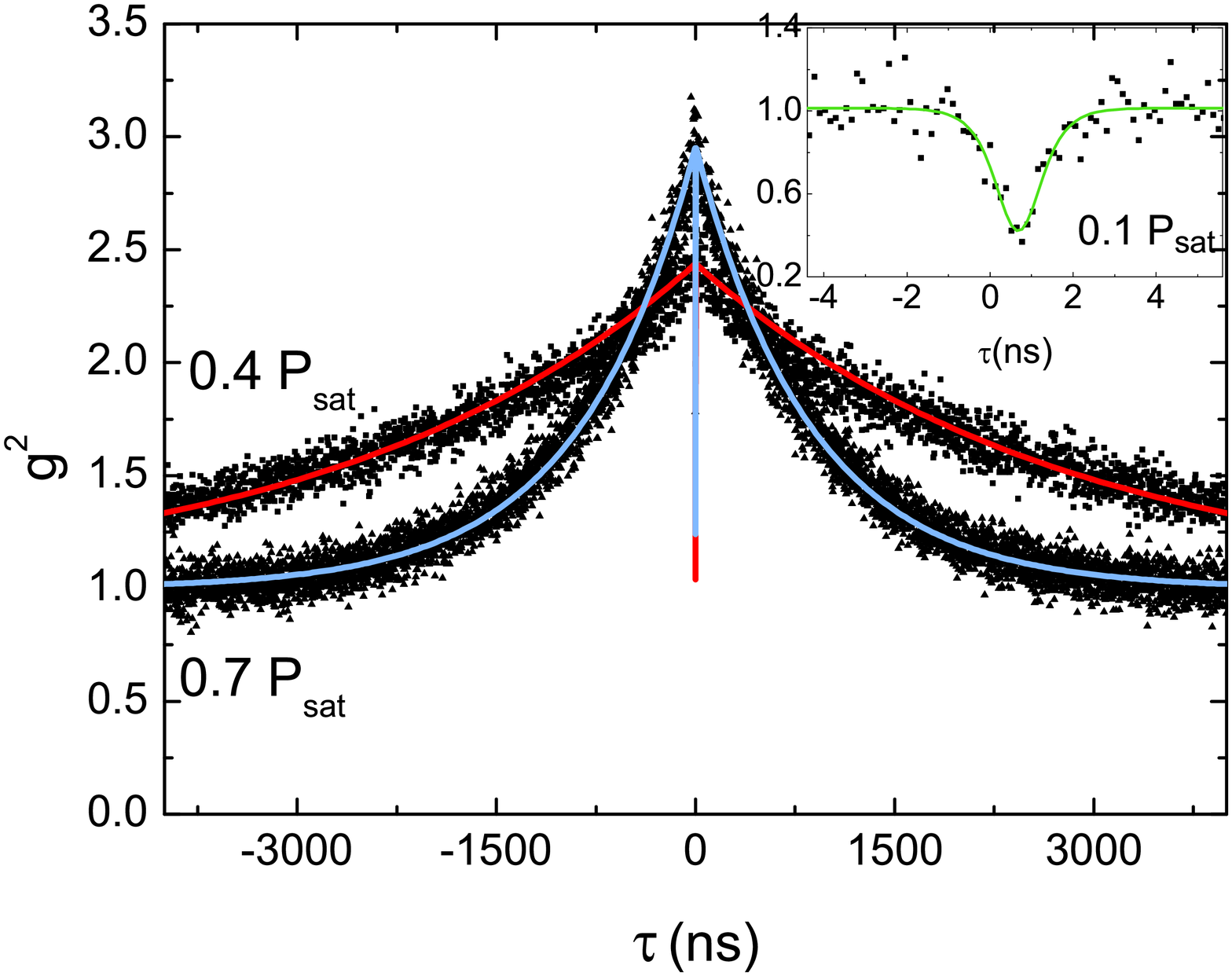}}
\subfigure[\label{g2em4}]{
\includegraphics[width=0.45\textwidth]{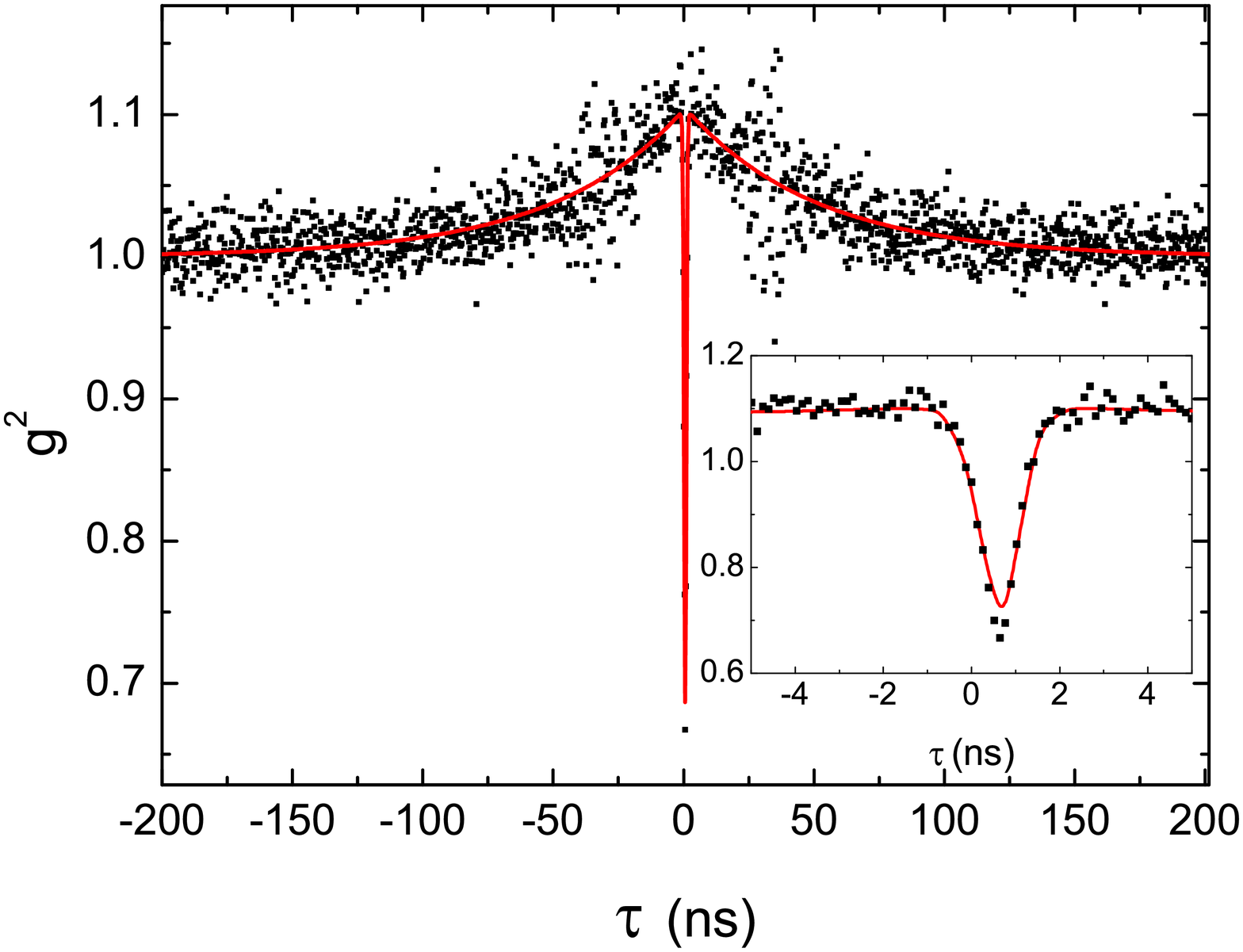}}
\caption{g$^2$ functions for emitters (2) and (4). (a) power dependent g$^2$ functions for emitter (2). The inset displays the antibunching dip at 0.1 P$_{sat}$ in detail. The fit confirms that the residual g$^2(0)$ is due to the timing jitter and the short $\tau_1=0.45$ ns constant only, confirming single emitter behaviour. (b) g$^2$ function of emitter (4) at $\textrm{P}=0.15$ $\textrm{P}_{sat}$. The inset shows the antibunching dip in detail, again the residual g$^2$ is governed by the jitter only confirming single emitter behaviour with $\tau_1=0.21$ ns.      \label{em2em4}}
\end{center}
\end{figure}

To prove the single emitter behaviour and to analyze the population dynamics of the investigated colour centres measurements of the g$^2$ function for different excitation powers have been performed. The results are displayed in figures \ref{em1dynamics}, \ref{em3dynamics} and \ref{em2em4}. As a signature of single emitters all curves show a pronounced antibunching dip at zero delay. Additionally, the measured g$^2$ functions exceed values of one for longer time scales giving rise to a bunching behaviour. The bunching signature is intensity dependent, becoming more pronounced for higher excitation powers: for emitter (1) g$^2$ values exceeding 10 are measured. The measurements also reveal differences in this bunching behaviour for the emitters: while for emitter (1) bunching becomes only visible for excitation well above saturation all other emitters exhibit photon bunching already for lower power. The time constants governing the bunching behaviour show a large variation among the emitters: the g$^2$ function of emitter (4) decays to a value of one for time delays of less than 50 ns even at low power, while for emitters (1), (2) and (3) time constants of more than 1 $\mu$s are observed for low power, decreasing to below 100 ns for high excitation power.\\
\begin{figure}[h]
\begin{center}
\includegraphics[width=0.3\textwidth]{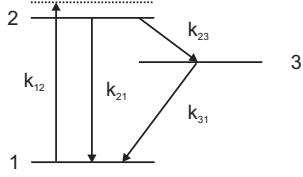}
\caption{Schematic representation of the three-level model employed to explain the population dynamics of the colour centres. \label{3levelscheme}}
\end{center}
\end{figure}
To explain this behaviour a three level system including a 'shelving-state' has to be considered. As the nature of the participating states is unknown, we use in a first approach a simplified model depicted in figure \ref{3levelscheme}: we assume that the excitation rate of level 2 depends linearly on the excitation power P: $k_{12}=\sigma$P. Level 1 and 2 are coupled via a fast radiative transition with the rate coefficient $k_{21}$, whereas level 3 acts as a 'shelving state' populated via the rate coefficient $k_{23}$ with the possibility of relaxation into the ground state via $k_{31}$. As long as the emitter resides in state 3 no photons on the radiative transition 2$\rightarrow$1 are detected. This simple model has been successfully employed to describe the dynamics of molecules involving shelving states \cite{Kitson1998}. Due to the very pronounced bunching behaviour at high powers we exclude efficient power dependent de-shelving processes as described in \cite{Wu2006,Aharonovich2010a} and thus assume in a first approach constant shelving and de-shelving rates $k_{23}$ and $k_{31}$, respectively. The g$^2$ function describes the dynamics and is defined by $\frac{n_2(\tau)}{n_2(\tau\rightarrow \infty)}$ resulting in:
\begin{equation}
g^{(2)}(\tau)=1-(1+a)\,e^{-|\tau|/\tau_1}+a\,e^{-|\tau|/\tau_2}
\label{g23level}
\end{equation}
The parameters $a$, $\tau_1$ and $\tau_2$ are given by  \cite{Wang2007b}:
\begin{eqnarray}
            \tau_{1,2}=2/(A\pm\sqrt{A^2-4B})\\[0.01 \textwidth]
            A=k_{12}+k_{21}+k_{23}+k_{31}\\[0.01 \textwidth]
            B=k_{12}k_{23}+k_{12}k_{31}+k_{21}k_{31}+k_{23}k_{31}\\[0.01 \textwidth]
            \label{apar} a=\frac{1-\tau_2k_{31}}{k_{31}(\tau_2-\tau_1)}
\end{eqnarray}
To describe the experimental results correctly we have to account for the timing jitter of the APDs used. To this end we convolute equation (\ref{g23level}) with the measured Gaussian response function of our setup ($\frac{1}{\sqrt{e}}$ width 354 ps). Figures \ref{em1dynamics}, \ref{em3dynamics} and \ref{em2em4} display the measured curves including the fits of equation (\ref{g23level}) convoluted with the response function. The deviation from the expected value g$^2(0)=0$ for a single emitter is fully explained by the timing jitter of the setup for low excitation power, thus unambiguously proving single emitter behaviour for the observed colour centres. Note that no correction for background luminescence was necessary to obtain these results, confirming negligible background contributions. We note that this also holds for higher excitation powers: For excitation powers of about P$_{sat}$ (emitters (1)-(3)) the deviation of the fitted $g^2(0)$ from the measured value is still only about 0.1. The results of the photon correlation measurements are thus consistent with values of $g^2(0)\le 0.1$.    \\
\begin{figure}
\begin{center}
\includegraphics[width=0.5\textwidth]{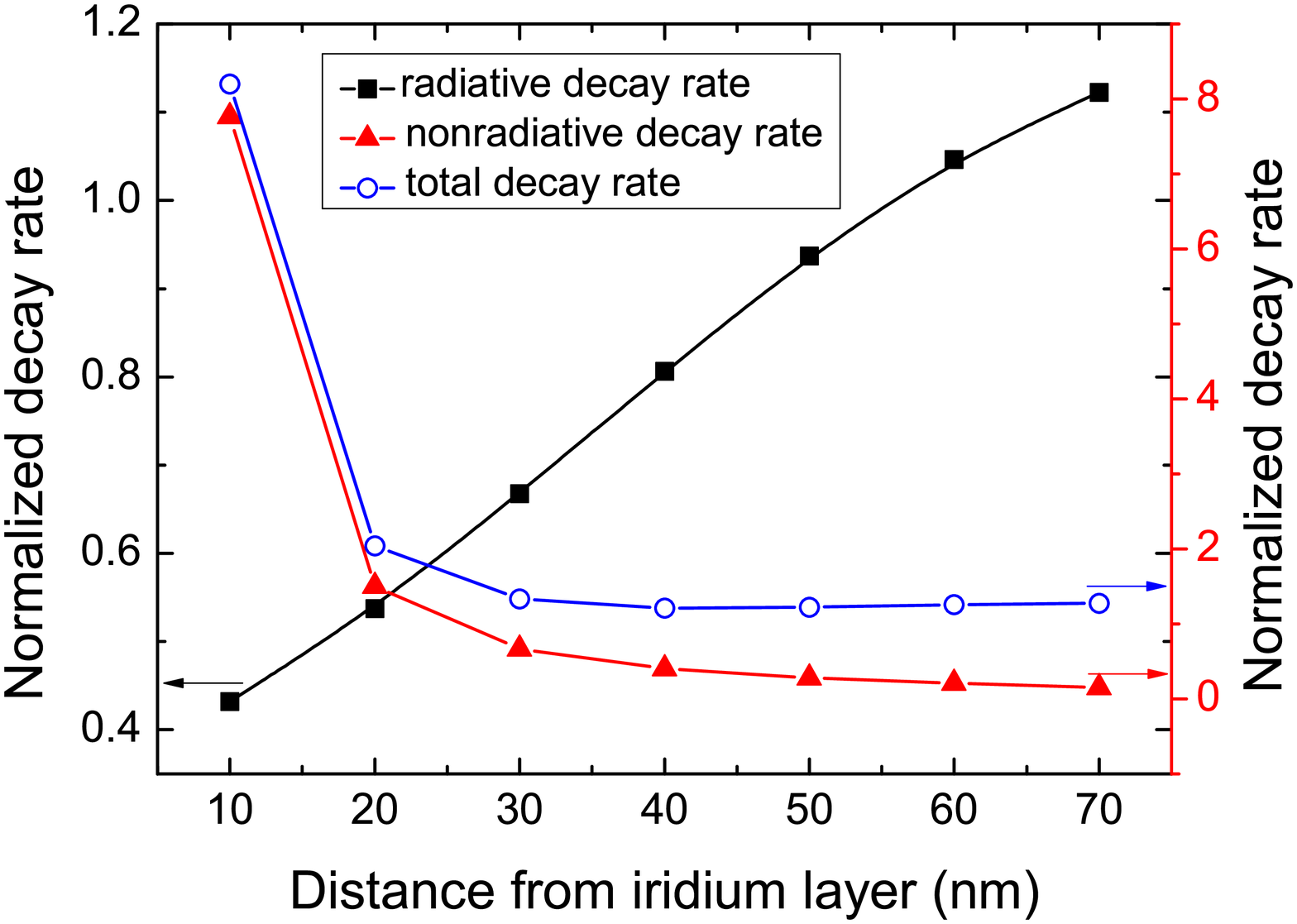}
\caption{Decay rates of an emitting dipole in diamond parallel to an iridium layer for different distances from the layer. The decay rates have been normalized to the rates of the dipole embedded in diamond without the presence of the iridium layer. \label{simulationresults}}
\end{center}
\end{figure}
The $\tau_1$ parameters at low excitation powers, indicating the lifetime of the excited state under the assumption $k_{21}\gg k_{23}+k_{31}$, which is valid in our case as shown later, show a large variation from 2 ns for emitter (3) down to 0.2 ns for emitter (4). The variation of the observed excited state lifetimes might have several reasons. First, the SiV-centre is assumed to possess a strong non-radiative decay channel \cite{Turukhin1996}. This non-radiative decay channel varies with local strain in different nano-diamonds \cite{Feng1993}. Second, the excited state lifetime is modified by the presence of the metal layer.  To estimate this influence we perform three-dimensional finite-difference time domain calculations (FDTD Solutions, Lumerical). To simplify the problem we assume an emitting dipole in a semi-infinite diamond slab (refractive index $\textrm{n}=2.4$) above a semi-infinite iridium layer ($\epsilon=-18+i\cdot25$ at 740 nm \cite{Weaver1977} ). As the orientation of the emitting dipole is unknown we simulate a dipole parallel to the iridium layer as a first estimate. The influence of the metal layer can be divided into two regimes: at distances $\gtrsim \frac {\lambda}{2}$ ($\approx$155 nm in diamond for the investigated colour centres) the lifetime usually shows an oscillatory behaviour due to far-field interaction of the dipole with its mirror image \cite{Buchler2005,Lukosz1977,Vion2010}. We disregard this regime as the diamond nano-crystals have a mean size of 130 nm and the emitter-metal-distance thus is $\lesssim$ 130 nm. In figure \ref{simulationresults} we present results for the radiative and non-radiative as well as the total decay rates for a dipole parallel to the iridium layer at distances between 10 nm and 70 nm from the iridium surface. Note that the rates are normalized with respect to the decay rates of an equivalent dipole in bulk diamond. At these distances there is a strong influence of the metal as the dipole near-field components efficiently couple to evanescent waves in the metal. The modification of the decay rates is especially pronounced for distances of less than 20 nm: due to the near-field coupling to the absorbing metal the non-radiative decay rate is enhanced by a factor of eight, while the radiative decay is simultaneously lowered by more than a factor of two. For a distance of 60 nm the decay rates roughly match the rates for the dipole without the presence of the iridium layer. Thus the presence of the metal layer might account for a spread of approximately a factor of eight in the measured lifetimes due to the varying positions of the colour centres in the nano-diamonds. However, we note that the lifetime reduction does not lead to an enhanced radiative decay due to the quenching by the metal in accordance with previous observations \cite{Buchler2005}.\\

The rate coefficients $k_{12}$, $k_{21}$, $k_{23}$, $k_{31}$ governing the population dynamics of the colour centres, are derived from the limiting values of the fitted parameters $a$, $\tau_1$ and $\tau_2$:
 \begin{eqnarray}
           \label{limv1} k_{31}=\frac{1}{(1+a^{\infty})\tau_2^{\infty}}\\[0.01 \textwidth]
           \label{limv2} k_{23}=k_{31}a^{\infty}\\[0.01 \textwidth]
           \label{limv3} k_{21}=\frac{1}{\tau_1^0}-k_{23} \qquad \textrm{for} \qquad k_{21}+k_{23}>k_{31}
\end{eqnarray}
where the superscript ${}^{\infty}$ denotes the limit for high excitation power and ${}^{0}$ denotes the limit for vanishing excitation power. The values of $a$, $\tau_1$ and $\tau_2$ for emitters (1) and (3) are displayed in figures \ref{konstem1} and \ref{konstem3}. The data indicate the limiting values of $a$, $\tau_1$ and $\tau_2$ despite the fact that some oscillatory behaviour is observed. This might be due to experimental reasons: above saturation spatial drifts changing the intensity impinging onto the emitter do no longer lead to measurable photon count rate changes, thus making detection of drifts and ensuring of constant excitation intensity very challenging. Nevertheless, a rough estimate of the rate coefficients is possible for emitters (1) and (3): we obtain $k_{21}=564$ MHz, $k_{23}=1.4$ MHz, $k_{31}=0.14$ MHz for emitter (1) and $k_{21}=469$ MHz, $k_{23}=6.7$ MHz, $k_{31}=5.0$ MHz for emitter (3). With these rate coefficients and the measured saturation power I$_{sat}$ we obtain the power dependence of $a$, $\tau_1$ and $\tau_2$ shown as solid lines in figures \ref{konstem1} and \ref{konstem3}. The curves qualitatively resemble the observed behaviour with deviations especially for $\tau_2$ and $a$ at low powers.\\
 In the following we propose an extension of the simple three level model in figure \ref{3levelscheme} allowing to account for these deviations. For the discussion we exemplarily use the data obtained from emitter (3). As an alternative to the limiting values used in equation (\ref{limv1}) we now employ the limiting value of $\tau_2$ for zero excitation:
\begin{equation}
k_{31}=\frac{1}{\tau_2^0} \label{tau20}
\end{equation}By using equation (\ref{tau20}) instead of (\ref{limv1}) together with equation (\ref{limv2}) and (\ref{limv3}), we obtain an 'alternative' set of rate parameters: $k_{21}=465$ MHz, $k_{23}=11.3$ MHz, $k_{31}=0.49$ MHz. Thus the value of $k_{31}$ seems to be inconsistent for high and low excitation power, indicating an intensity dependent de-shelving process. As mentioned before we exclude a de-shelving process linearly dependent on excitation power \cite{Wu2006,Aharonovich2010a,Fleury2000}, as this model would lead to
\begin{eqnarray}
\tau_2^{\infty}=0 \qquad
a^{\infty}=0
\end{eqnarray} which is not consistent with our observations. We therefore tentatively modify the intensity dependence of the rate $k_{31}$ to follow a saturation law, including an intensity independent rate $k_{31}^0$:
\begin{equation}
k_{31}=\frac{d\cdot \textrm{P}}{\textrm{P}+c}+k_{31}^0
\end{equation} The introduction of a saturating de-shelving process is further motivated by the possible explanation of this process as an excitation from the shelving state to higher lying states, as e.g. found for molecules in \cite{Fleury2000}. This excitation process might intrinsically exhibit a saturation behaviour. For the new model we calculate $k_{23}$, $k_{21}$, $k_{31}^0$, $d$ under the assumption $k_{21}+ k_{23}> k_{31}^0$:
\begin{eqnarray}
\label{limnew1} k_{31}^0=\frac{1}{\tau_2^0}\\
d=\frac{\frac{1}{\tau_2^{\infty}}-(1+a^{\infty})\frac{1}{\tau_2^0}}{a^{\infty}+1}\\
k_{23}=\frac{1}{\tau_2^{\infty}}-k_{31}^0-d\\
\label{limnew4} k_{21}=\frac{1}{\tau_1^0}-k_{23} \label{limnew4}
\end{eqnarray}
For emitter (3) we obtain with equations (\ref{limnew1}) to (\ref{limnew4}) and $a^{\infty}$, $\tau_1^0$, $\tau_2^0$, $\tau_2^{\infty}$ derived from the measurements: $k_{31}^0=0.50$ MHz, $k_{23}=6.72$ MHz, $k_{21}=469.48$ MHz and $d=4.55$ MHz. Now the parameter $c$ as well as the proportionality constant $\sigma$ for the excitation rate $k_{12}=\sigma P$, can be determined by fitting the intensity dependent data for $a$ with equation (\ref{apar}) ($\sigma$ is no longer accessible from the saturation measurement in this model). The results are shown in figure \ref{konstem3}, we obtain $c=98.1 \,\,\mu$W and $\sigma=5.72$ MHz$\slash \mu$W. With these values of $c$ and $\sigma$ the intensity dependence of $\tau_1$ and $\tau_2$ is also well described. We thus conclude that a complex power dependence of the de-shelving rate has to be accounted for to properly model the SiV-centre dynamics. To prove the general validity of this model, it has to be tested on a larger number of emitters.\\
The rate coefficients determine the maximum obtainable photon count rate I$_{\infty}$ for cw excitation and the maximum steady state population of the excited state $n_2^\infty=n_2(\tau\rightarrow \infty,k_{12}\rightarrow \infty)$:
\begin{equation}
\textrm{I}_{\infty}=\eta_{coll}\,\eta_{qe}\, k_{21}\, n_2^\infty=\eta_{coll}\, \eta_{qe}\,\frac{k_{21}}{1+\frac{k_{23}}{k_{31}^0+d}}
\end{equation}
For the simple three level model $k_{31}^0+d$ has to be replaced by the intensity independent rate $k_{31}$. As the orientation of the emitting dipole inside the nano-crystal is unknown the actual collection efficiency $\eta_{coll}$ includes an uncertainty of a factor of 2.4 as discussed in section \ref{photoncount}. In addition, the rate coefficients have to be considered as a rough estimate, thus determining the quantum efficiency $\eta_{qe}$ from the measured photon rate is not reliable. Nevertheless, by using the rate coefficients we illustrate the influence of the shelving state on the maximum photon count rate for cw excitation: for an off-resonantly pumped two-level-system with decay rate $k_{21}$ we expect a maximum photon rate
\begin{equation}
\textrm{I}_{\infty}=\eta_{coll}\, \eta_{qe}\, k_{21} n_2^\infty=\eta_{coll}\, \eta_{qe}\,k_{21}
\end{equation}
Thus we deduce that for emitter (3) the presence of the shelving state changes the maximum photon rate only by about a factor of two as $n_2^\infty=0.43$ compared to $n_2^\infty=1$ for an off-resonantly pumped two level system. In contrary, for emitter (1) the presence of the shelving state leads to $n_2^\infty=0.09$ thus potentially lowering the maximum photon rate by a factor of more than 10. As the nature of the shelving state and the relaxation process is unknown, the reason for the differences in bunching behaviour remains unclear. We emphasize that the large spread in excited state lifetimes in combination with large differences in shelving state dynamics suffice to explain large variations in photon emission rates of SiV-centres.
\subsection{ Single emitter spectroscopy at cryogenic temperatures}
\begin{figure}[h!!!]
\begin{center}
\includegraphics[width=0.5\textwidth]{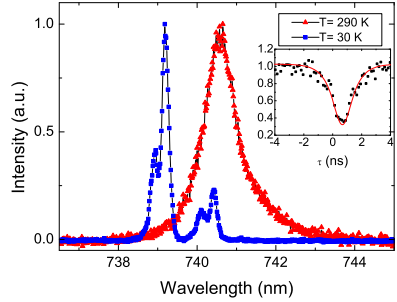}
\caption{Temperature dependent spectra of emitter (5). The spectra have been measured at 290 K (red triangles) and 30 K (blue squares). The inset displays the g$^2$ function of emitter (5) measured at 1.3 $\mu$W. The fit confirms that the residual g$^2$(0) value is fully explained by the timing jitter of the setup, thus proving existence of a single emitter. \label{em5kryo}}
\end{center}
\end{figure}
Spectroscopy of colour centres at cryogenic temperatures is an important issue for their application in quantum optics: underlying fine structures of ZPLs can only be observed at cryogenic temperatures reducing phonon-broadening; additionally the determination of low temperature line-widths is crucial for applications involving coherent processes. We characterize emitter (5) at room temperature and at 30 K. Figure \ref{em5kryo} displays the measured spectra. Note that g$^2$ measurements prove single emitter behaviour for both temperatures. Emitter (5) has a room temperature line-width of 1 nm. Upon cooling to 30 K a fine structure evolves consisting of four components at 740.42 nm, 740.11 nm, 739.19 nm and 738.91 nm. The line-width of the individual components reduces to 0.17 nm. This four line fine structure is characteristic for SiV-centres and has so far been observed only for ensembles of SiV-centres in single-crystals of high quality \cite{Clark1995,Sternschulte1994}, thus providing significant evidence for the identification of the observed colour centres as SiV-centres. The spacing of the outer components is 1.5 nm. For SiV-centres in single-crystals this spacing increases from 0.7 nm (no stress) to 2.4 nm by applying a stress of 0.5 GPa along 001-direction \cite{Sternschulte1994}. Thus the observed spacing can be explained by stress inside the nano-crystal. The spectrum shifts by 1.5 nm to shorter wavelengths when cooling to 30 K (highest fine structure peak compared to room temperature ZPL). This shift is also characteristic for SiV-centres in poly-crystalline CVD-diamonds: Ghorokovsky et al. \cite{Gorokhovsky1995} as well as Feng et al. \cite{Feng1993} observed a blue shift of 1.4 nm and 1.2 nm, respectively, when cooling to 100 K and no further shift for lower temperatures. For the observed single SiV-centre the line shifting also stops around 110 K, thus resembling the observations for ensembles.
\section{Conclusion}
We have identified an advanced material system for single photon sources in diamond employing high quality CVD-nano-diamonds on iridium produced via spin coating seeding and microwave CVD-growth. Due to impurities in the growth process single SiV-centres are formed. The iridium substrates together with the small volume nano-diamonds provide the major advantage of single photon emission with negligible contributions of background fluorescence. Furthermore, the SiV colour centres observed here are much better than their reputation and exhibit very promising behaviour as single photon sources: line-widths as low as 0.7 nm could be detected with a fraction of up to 88\% of photons emitted into the ZPL. In addition, the observed colour centres feature count rates up to 4.8 Mcps thus being the brightest diamond based single photon sources to date. By performing spectroscopy at cryogenic temperatures we verify the identification of these colour centres as SiV-centres, measuring for the first time the fine structure splitting of a single SiV-centre.
\section*{Acknowledgements}
The SEM-analysis of the samples was performed by J. Schmauch  (Universit\"at des Saarlandes, Saarbr\"u\-cken, Germany) and M. Fischer  (Universit\"at Augsburg, Germany). We acknowledge support with the spin-coating procedure by L. Marquant and valuable help by R. Albrecht  (Universit\"at des Saarlandes, Saarbr\"u\-cken, Germany). The SiV-structure was illustrated by C. Hepp, B. Weigand assisted with the g$^2$-fitting procedure (Universit\"at des Saarlandes, Saarbr\"u\-cken, Germany).
The project was financially supported by the Deutsche Forschungsgemeinschaft and the Bundesministerium f\"ur Bildung und Forschung (network EphQuaM, contract 01BL0903).

\end{document}